\documentclass[traditabstract,referee,pdftex]{aa} 
\usepackage{graphicx}
\usepackage{txfonts}
\begin{document}
   \title{Thermal Infrared Observations of Asteroid (99942) Apophis with Herschel\thanks{{\it Herschel} is an
          ESA space observatory with science instruments provided by European-led Principal
          Investigator consortia and with important participation from NASA.}}


   \author{
          T. G. M\"{u}ller \inst{1},
          C. Kiss \inst{2},
          P. Scheirich \inst{3},
          P. Pravec \inst{3},
          L. O'Rourke \inst{4},
          E. Vilenius \inst{1},
          \and
          B. Altieri \inst{4},
          }

   \institute{   {Max-Planck-Institut f\"{u}r extraterrestrische Physik,
                 Postfach 1312, Giessenbachstra{\ss}e,
                 85741 Garching, Germany
                 }
          \and
                 {Konkoly Observatory, Research Center for Astronomy and
                 Earth Sciences, Hungarian Academy of Sciences;
                 Konkoly Thege 15-17, H-1121 Budapest, Hungary
                 }
          \and
                 {Astronomical Institute, Academy of Sciences of the Czech Republic,
                 Fri\v{c}ova 1, CZ-25165 Ond\v{r}ejov, Czech Republic
                 }
          \and
                 {European Space Astronomy Centre (ESAC), European Space Agency,
                 Apartado de Correos 78, 28691 Villanueva de la Ca\~nada,
                 Madrid, Spain
                 }
          }

   \date{Received ; accepted }

\abstract{The near-Earth asteroid (99942) Apophis is a potentially hazardous asteroid.
	  We obtained far-infrared observations of this asteroid with the Herschel Space
          Observatory's PACS instrument at 70, 100, and 160\,$\mu$m. These were taken
          at two epochs in January and March 2013 during a close Earth encounter.
          These first thermal measurements of Apophis were taken at similar phase
          angles before and after opposition.
          We performed a detailed thermophysical model analysis 
          by using the spin and shape model recently derived from applying a 2-period
          Fourier series method to a large sample of well-calibrated photometric observations.
	  We find that the tumbling asteroid Apophis has an elongated shape with a
          mean diameter of 375$^{+14}_{-10}$\,m (of an equal volume sphere) and a
          geometric V-band albedo of 0.30$^{+0.05}_{-0.06}$.
	  We find a thermal inertia in the range 250-800\,Jm$^{-2}$s$^{-0.5}$K$^{-1}$
          (best solution at $\Gamma$ = 600\,Jm$^{-2}$s$^{-0.5}$K$^{-1}$),
          which can be explained by a mixture of low conductivity fine regolith with
          larger rocks and boulders of high thermal inertia on the surface. The thermal
          inertia, and other similarities with (25143)~Itokawa indicate that Apophis
          might also have a rubble-pile structure. If we combine the new size value with
          the assumption of an Itokawa-like density and porosity we estimate a mass between
          4.4 and 6.2 $\cdot$ 10$^{10}$\,kg which is more than 2-3 times larger
          than previous estimates. We expect that the newly derived properties will
          influence impact scenario studies and influence the long-term orbit predictions
          of Apophis.
	  }


\authorrunning{M\"uller et al.}
\titlerunning{Herschel observations of (99942) Apophis}

   \maketitle
%

\section{Introduction}

The near-Earth asteroid 99942~Apophis was discovered in 2004 (Minor Planet Supplement 109613)
and found to be on an Aten-type orbit\footnote{The current orbit's perihelion is at 0.746\,AU,
aphelion at 1.0985\,AU, with a=0.922\,AU, i=3.33$^{\circ}$, e=0.191.} crossing the Earth's orbit
in regular intervals. At that time, the object raised serious concerns following the discovery
that it had a 2.7\% chance of striking the planet Earth in 2029\footnote{\tt
http://neo.jpl.nasa.gov/risk \\
http://newton.dm.unipi.it/neodys}.
Immediate follow-up observations to address these concerns took
place and provided predictions that
eliminated the possibility of collision in 2029, although it does
enter below the orbit of the geostationary satellites at that time. 
However there did remain the possibility of Apophis passing through
a precise region in space (gravitational keyhole) which could set
it up for an impact in the mid-term future (Farnocchia et al.\ \cite{farnocchia13}).
Apophis remains an object with one of the highest statistical
chances of impacting the Earth among all known near-Earth Asteroids.

The studies performed to determine the impact probability require
a clear set of physical properties in order to understand the
orbital evolution of this asteroid (\v{Z}i\v{z}ka \& Vokrouhlick\'y \cite{zizka11};
Farnocchia et al.\ \cite{farnocchia13}; Wlodarczyk \cite{wlodarczyk13}).
The lack of availability of such properties (albedo, size, shape, rotation,
physical structure, thermal properties) is a major limiting factor
which leads to uncertainties in the role played by non-gravitational effects
on that orbit. The Yarkovsky effect due to the recoil of thermally
re-radiated sunlight is the most important of these non-gravitational
effects.

Besides the input to the orbit evolution, the physical properties
serve also to address the possible implications if an impact were
to occur. A solid body of 300\,m versus a rubble pile hitting the
Earth implies different levels of severity as regards its ability
to pass through the atmosphere unscathed to create regional versus
grandscale damage.

Delbo et al.\ (\cite{delbo07a}) determined from polarimetric observations
an albedo of 0.33 $\pm$ 0.08 and an absolute magnitude of H = 19.7 $\pm$ 0.4\,mag.
These values led to a diameter of 270 $\pm$ 60\,m, slightly
smaller than earlier estimates in the range 320 to 970\,m, depending
on the assumed albedo.
Binzel et al.\ (\cite{binzel09}) described the results of
observations they performed in the visible to near infrared
(0.55 to 2.45\,$\mu$m) of Apophis where they compared and modeled
its reflectance spectrum with respect to the spectral and
mineralogical characteristics of likely meteorite analogs.
Apophis was found to be an Sq-class asteroid that most closely
resembled LL ordinary chondrite meteorites in terms of spectral
characteristics and interpreted olivine and pyroxene abundances.
They found that composition and size similarities of Apophis
with (25143) Itokawa suggested a total porosity of 40\% as a
current best guess for Apophis. Applying these parameters to
Apophis yielded a mass estimate of 2 $\cdot$ 10$^{10}$\,kg with
a corresponding energy estimate of 375 Megatonnes (Mt) TNT for
its potential hazard. Substantial unknowns, most notably the
total porosity, allowed uncertainties in these mass and energy
estimates to be as large as factors of two or three.

Up to the time of our own observations, there were no thermal infrared measurements existing
on this asteroid. Observations from the Spitzer Space Telescope were not possible
as Apophis was not in the Spitzer visibility region during the
remainder of its mission. Due to the fact that there was no close
encounter with Earth between discovery and now, there are 
also no groundbased N-/Q-band observations, no Akari and
also no WISE observations available.

We observed this near-Earth asteroid with the Herschel Space Observatory's (Pilbratt et al.\ \cite{pilbratt10}) PACS (Photodetector Array Camera and Spectrometer) instrument (Poglitsch et al.\ \cite{poglitsch10}) at far-infrared wavelengths (Section \ref{sec:obs}). We present our thermophysical
model (TPM) analysis (Section \ref{sec:tpm}) and discuss the results (Section
\ref{sec:dis}).

\section{Far-infrared observations with Herschel-PACS}
\label{sec:obs}

\begin{figure}[h!tb]
 \rotatebox{0}{\resizebox{8cm}{!}{\includegraphics{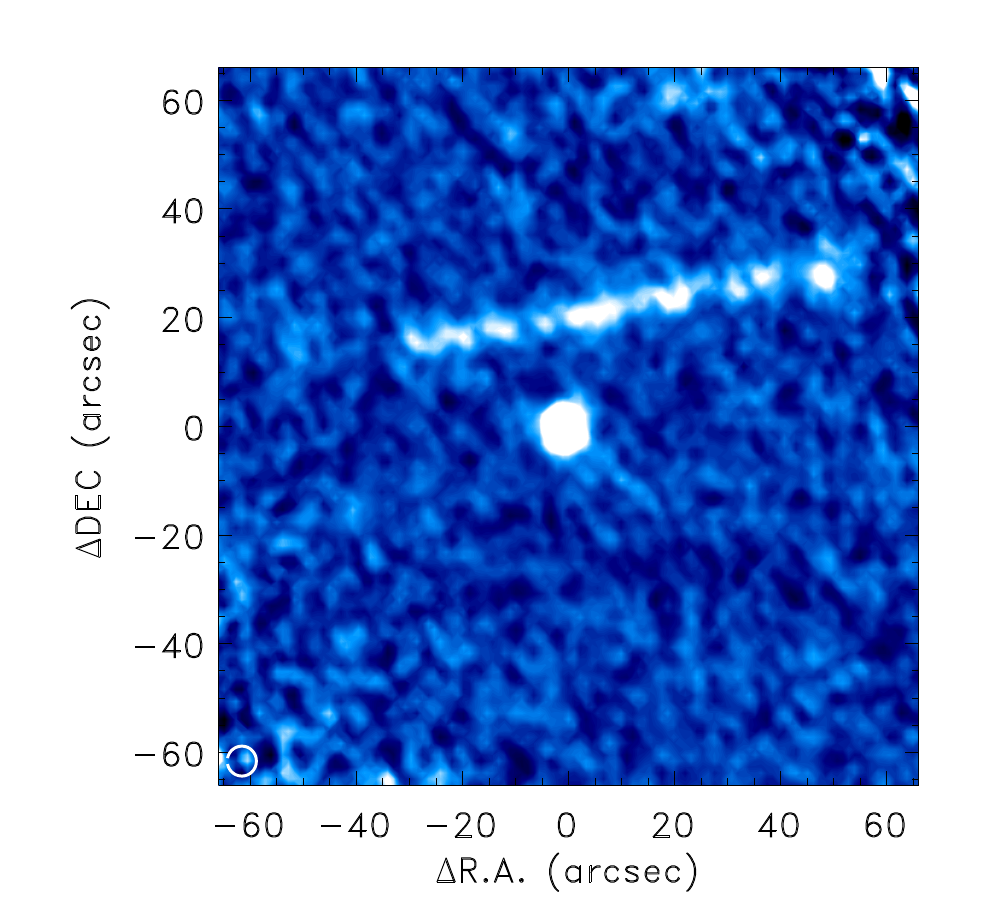}}}
 \rotatebox{0}{\resizebox{8cm}{!}{\includegraphics{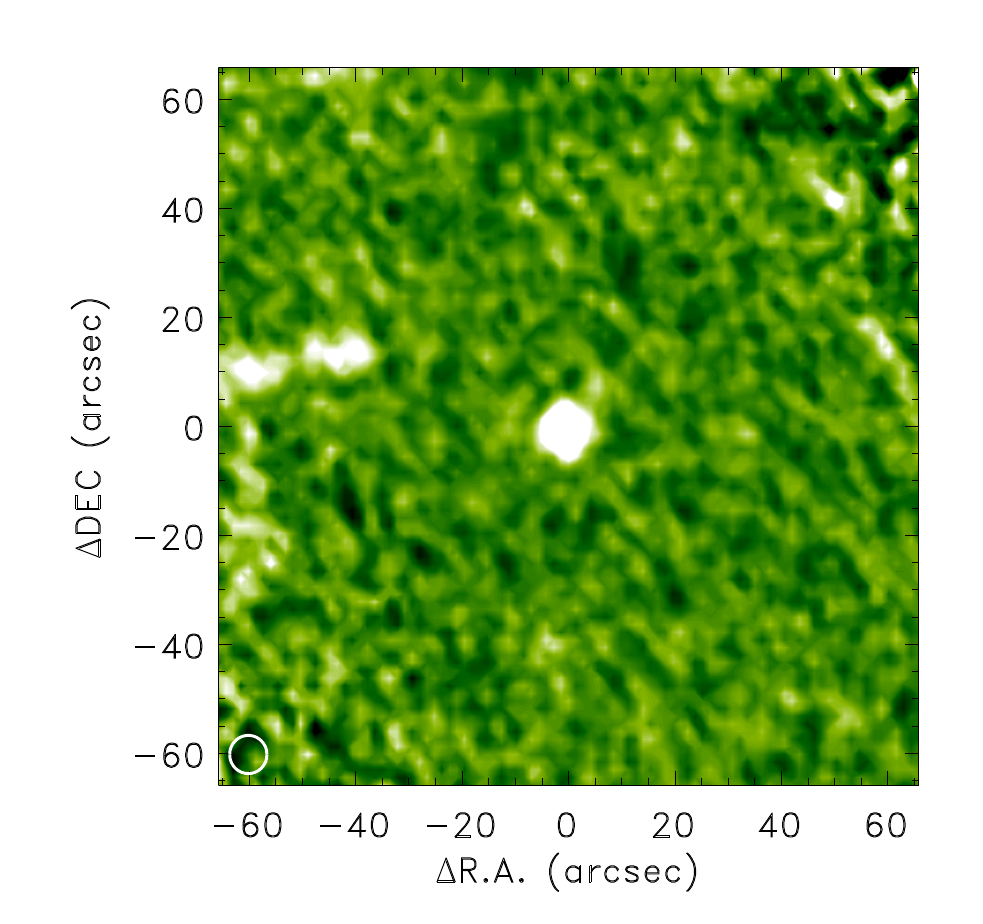}}}
 \rotatebox{0}{\resizebox{8cm}{!}{\includegraphics{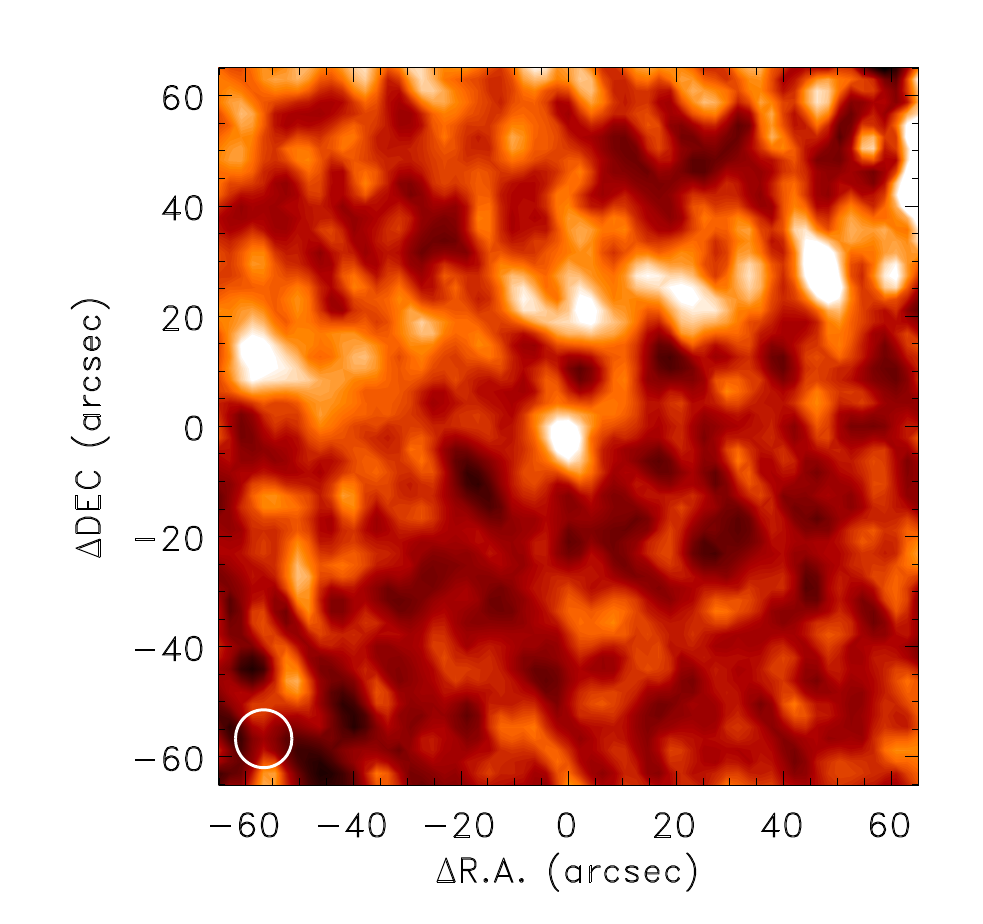}}}
  \caption{The object-centered images of the target in the 3 PACS filters
           for the first visit on Jan.\ 6, 2013.  Top: blue (70\,$\mu$m),
           middle: green (100\,$\mu$m), bottom: red (160\,$\mu$m).
     \label{fig:pacs}}
\end{figure}

The far-infrared observations with the Herschel Space Observatory
were performed in several standard PACS mini scan-map observations
in tracking mode. The observations took place on Jan.\ 6, 2013
(four individual observations)
and on Mar.\ 14, 2013 (one individual observation).
Each individual observation consisted of several repetitions of a mini
scan-map. The observational circumstances are listed in Table~\ref{tbl:obspacs}.
During the first epoch all three PACS
filters at 70 (blue), 100 (green), and 160\,$\mu$m (red band) were used,
while in the second epoch we concentrated only on the 70/160\,$\mu$m filter setting
due to observing time limitations.
Each measurement consisted of a mini scan-map with 10 scan-legs of 3\,arcmin
length and separated by 4\,arcsec, the scan direction was 70$^{\circ}$ (along
the diagonal of the detector arrays), and the scan-speed was 20$^{\prime \prime}$/s.
Each scan-leg is centered on the true object position at scan mid-time.
The PACS photometer takes data frames with 40\,Hz, but binned onboard
by a factor of 4 before downlink. The total duration of our Herschel-PACS
observations was about 2\,h during the first epoch, split in 4 measurements
of about 30\,min each: 2$\times$ 6 map repetitions in the blue, 2$\times$ 7
map repetitions in the green band, each time with the red channel in parallel.
During the second visit we only executed one single measurement of about 1.4\,h
which corresponds to 18 map repetitions in the blue/red filter setting. In this case
we split the data into 6 individual datasets with 3 repetitions each.

Figure~\ref{fig:pacs} shows the object-centered images of the first visit in
January 2013.
They are produced by stacking all frames of a given band on the source position
in the first frame.
The background structures in these figures are not real and related to 
background source artefacts caused by the re-centering of images on the
rapidly changing Apophis position. During the first visit Apophis was moving
in a clean part of the sky without any significant sources along the object's
path. During the second visit the source moved over faint objects located
in a field of diffuse background emission which we could not entirely
eliminate in the reduction process. We followed the object's flux (in the
background-subtracted images) and noticed a 1-2\,mJy residual background
emission in parts of the object's trajectory (see footnote in Table~\ref{tbl:obspacs}).
In addition to the six sub-images we also combined all background-free and
clean images (repetitions 4-9, 16-18) to obtain a final object-centered
map for high-quality photometry.

We performed aperture photometry on the final calibrated images and
estimated the flux error via photometry on artificially implemented
sources in the clean vicinity around our target.
The fluxes were finally corrected for colour terms due to the differences
in spectral energy distribution between (99942)~Apophis and the assumed
constant energy spectrum $\nu$~F$_{\nu}$ = const.\ in the PACS
calibration scheme. The calculated colour-corrections for our best
Apophis model solution are 1.005, 1.023, 1.062, at 70.0, 100.0, and
160.0\,$\mu$m respectively. These values agree with the expected
corrections\footnote{PACS technical report PICC-ME-TN-038, v1.0:
{\tt http://herschel.esac.esa.int/\-twiki/\-pub/\-Public/\-PacsCalibrationWeb/\-cc\_report\_v1.pdf}}
for objects with temperatures around 250\,K.
The absolute flux calibration error is 5\% in all three bands. This error is
based on the model uncertainties of the fiducial stars used in the
PACS photometer flux calibration scheme (Nielbock et al.\ \cite{nielbock13};
Balog et al.\ \cite{balog14}). Since this error is identical for all our
observations, we consider it at a later stage in the discussion about the
quality of our derived properties.
The final monochromatic flux densities and their
flux errors at the PACS reference wavelengths 70.0, 100.0 and 160.0\,$\mu$m
are listed in Table~\ref{tbl:obspacs}.

\section{Radiometric analysis}
\label{sec:tpm}

\subsection{Shape and spin properties}
\label{sec:shape_sv}

\begin{figure}[h!tb]
  \rotatebox{90}{\resizebox{!}{\hsize}{\includegraphics{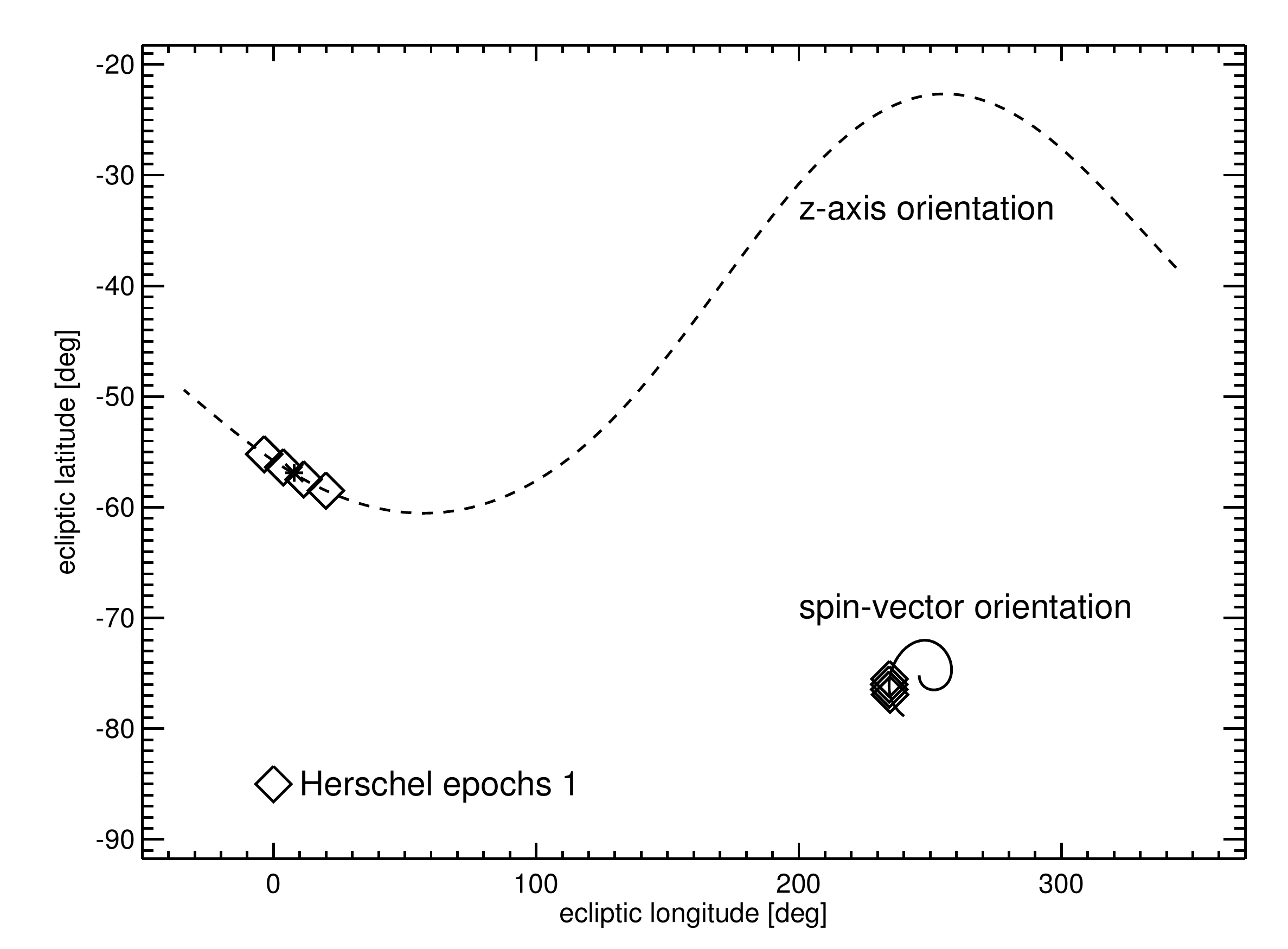}}}
  \rotatebox{90}{\resizebox{!}{\hsize}{\includegraphics{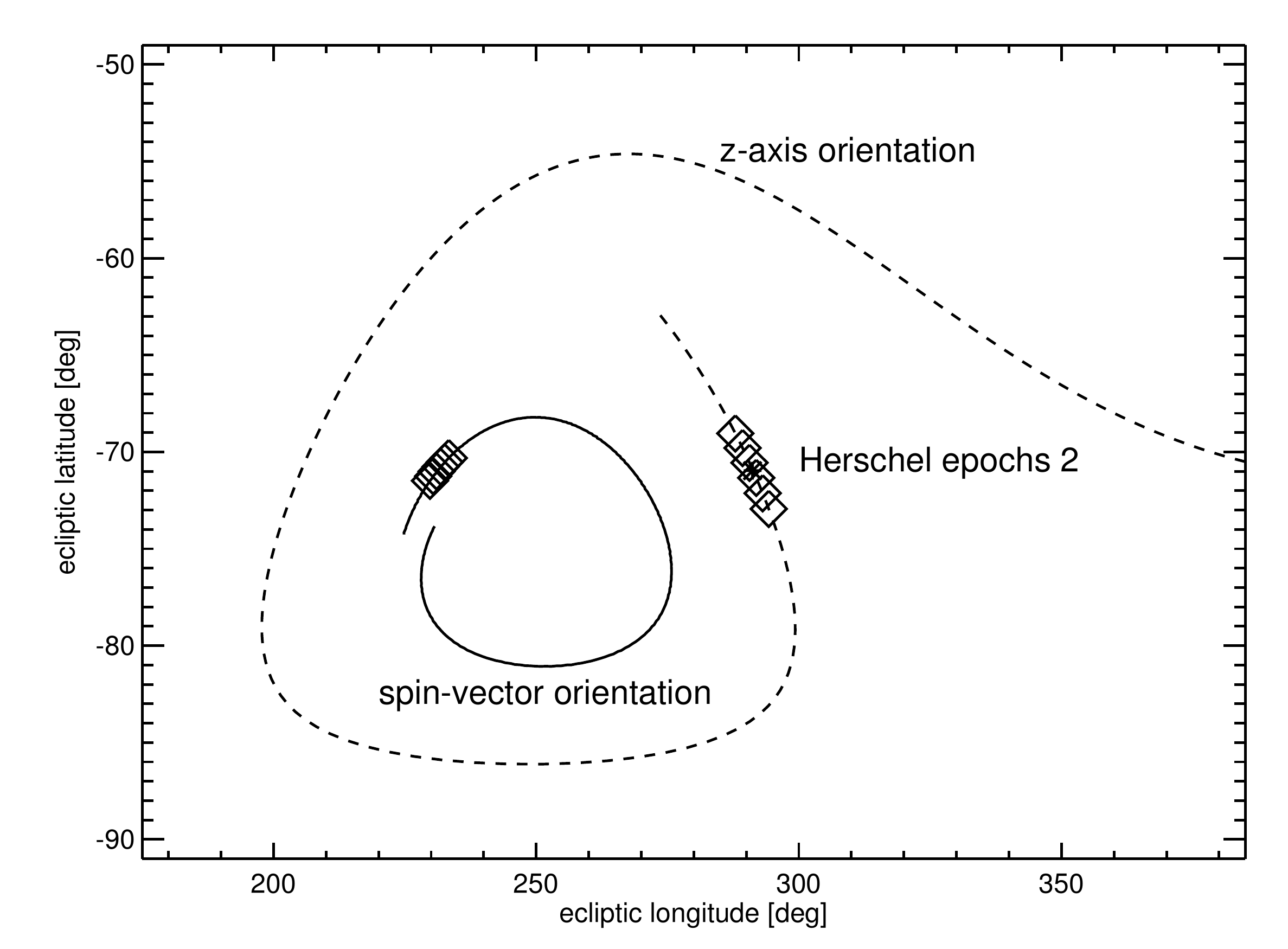}}}
  \caption{Variations of the object's z-axis and spin-vector orientation during
           a full rotation of 30.56\,h, starting about 26\,h before the first Herschel
           measurement and ending about 2\,h after the last measurement.
           The epochs of the Herschel observations (from Table~\ref{tbl:obspacs}) are
           shown as diamonts.
           The z axis (dashed line) is connected to the largest moment of
           inertia in the asteroid's co-rotating coordinate frame, the solid
           line shows how the orientation of the object's spin axis changes
           with time. Top: covering the first observations on Jan.\ 6, 2013;
           bottom: covering the second observation on Mar.\ 14, 2013.
     \label{fig:tumbler}}
\end{figure}


Pravec et al.\ (\cite{pravec14}) found that Apophis has
a non-principal axis rotation and it is in a moderately
excited Short Axis Mode state. The strongest observed
lightcurve amplitude\footnote{The full (peak-to-trough) amplitude
of the strongest lightcurve frequency
$2 P_{1}^{-1} = 2 (P_{\phi}^{-1} - P_{\psi}^{-1})$ is
0.59 $\pm$ 0.03\,mag, with the precession period P$_{\phi}$ = 27.38 $\pm$ 0.07\,h and the
rotation period P$_{\psi}$ = 263 $\pm$ 6\,h.} is related to a retrograde rotation with
P$_1$ = 30.56 $\pm$ 0.01\,h and the angular momentum
vector at ($\lambda_{ecl}$, $\beta_{ecl}$) = (250$^{\circ}$, -75$^{\circ}$).
The relevant parameters for our radiometric analysis are (i) the
orientation of the object at the time of the Herschel observations,
given by the object's z axis which is connected to the largest
moment of inertia in the asteroid's co-rotating coordinate frame,
and the angle $\phi_0$ which specifies the rotation angle of the body at
the given julian date.
(ii) the rotation history of the object to account for thermal 
inertia effects (the thermal inertia is responsible for "transporting"
heat to the non-illuminated parts of the surface).

\begin{figure*}[h!tb]
  \resizebox{20cm}{!}{\includegraphics{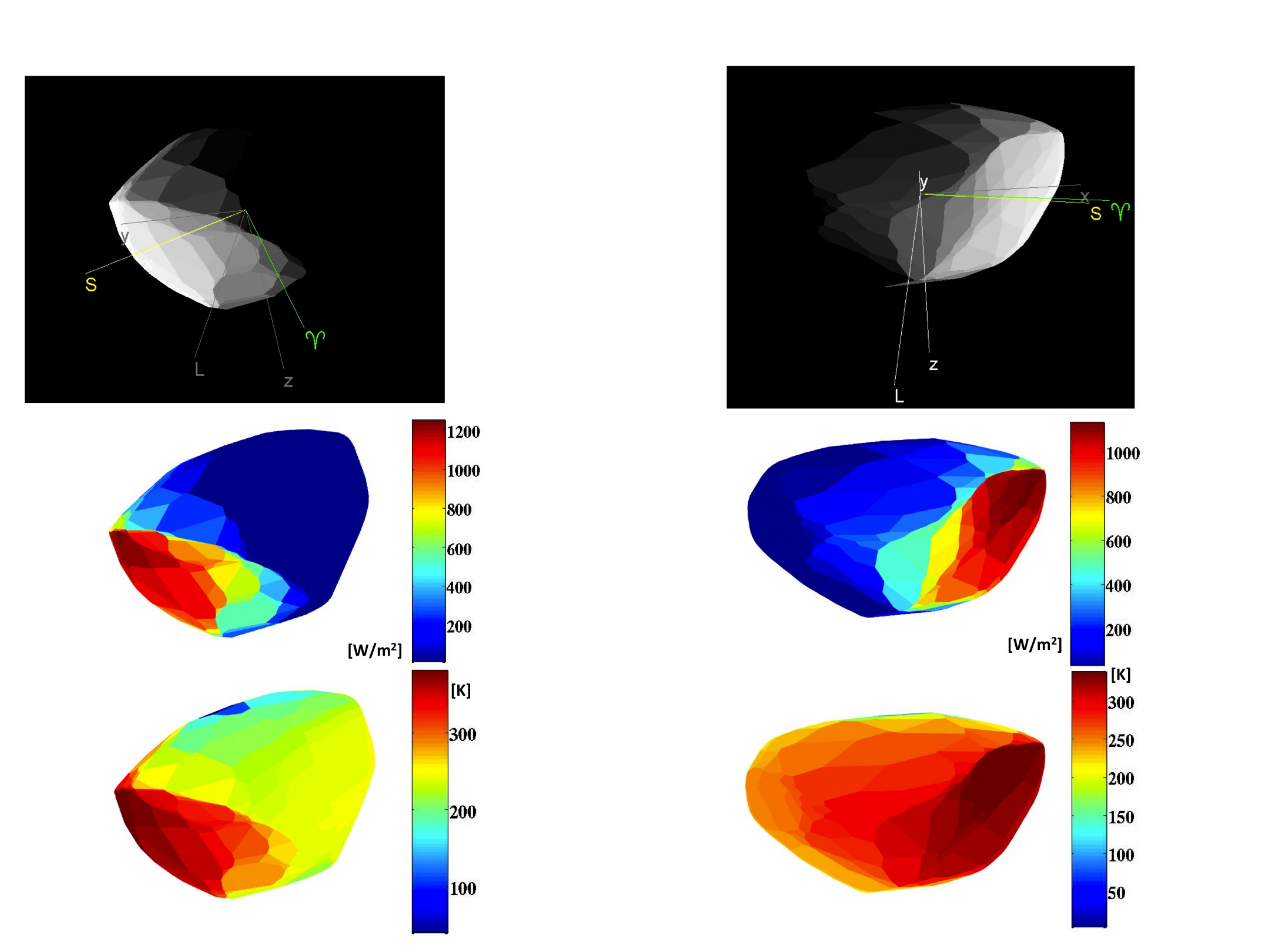}}
  \caption{Viewing geometry during the two Herschel observing epochs at phase
        angles of roughly 60$^{\circ}$ angle before (left)
	and after opposition (right).
	Top: calculated observing geometry on basis of the nominal solution in Pravec et al.\
	(\cite{pravec14}). L is fixed vector of angular momentum, the Aries sign is the
	X axis of the ecliptical frame, S is a direction to the Sun, and x, y, z are
	the axes of the asteroid co-rotating coordinate frame (corresponding to the
	smallest, intermediate and the largest moment of inertia of the body,
	respectively). Middle: The solar insolation in [W/m$^2$]. Bottom: TPM temperature
	calculations assuming a Itokawa-like thermal inertia of 600\,Jm$^{-2}$s$^{-0.5}$K$^{-1}$.
     \label{fig:tpm}}
\end{figure*}

Pravec et al.\ (\cite{pravec14}) were also able to reconstruct the 
physical shape model (see Figure~\ref{fig:tpm}) of Apophis following the work
by Kaasalainen (\cite{kaasalainen01}; \cite{kaasalainen01a}) and Scheirich et al.\ (\cite{scheirich10}).
The convex shape model with the non-principal axis rotation was determined by Pravec et al. to be the best-fit solution to the observed lightcurves from December 2012 to April 2013.
The available photometric observations were found to cover our Herschel measurements
in January 2013 very well. In March 2013 the situation is less
favorable and the photometric points are sparsely distributed
in the days before and after the Herschel observations
(see Fig.~6 in Pravec et al.\ \cite{pravec14}).

The lightcurve-derived shape model does not have absolute size information.
A dark (low albedo) and large object could explain the observed lightcurves
equally well as a bright (high albedo) but much smaller object.
For our analysis we used the physical shape model and the rotational
properties presented in Pravec et al.\ (\cite{pravec14}), the relevant
coordinates and angles connected to our thermal measurements
are listed in Table~\ref{tbl:obj_coord} and shown in the context of
a full rotation in Figure~\ref{fig:tumbler}.

\subsection{Thermophysical model analysis}

\begin{table*}
     \caption{Observing geometries (Herschel-centric) and final calibrated flux densities (FD).
      r$_{helio}$ is the heliocentric distance, $\Delta_{obs}$ the object's distance from
      Herschel, and  $\alpha$ is the phase-angle, with negative values after opposition.
      OD is Herschel's operational day, OBSID: Herschel's observation identifier.
      The repetitions specify the number of scan-maps
      performed and/or used to derived the given flux and error. The Herschel-centric
      apparent motions of Apophis were 205$^{\prime \prime}$/h (first visit in January 2013)
      and 58$^{\prime \prime}$/h (second visit in March 2013).
     \label{tbl:obspacs}}
     \begin{tabular}{lrrrrrrlrr}
        \noalign{\smallskip}
        \hline
        \hline
        \noalign{\smallskip}
Julian Date & $\lambda_{ref}$ & FD    & FD$_{err}$   & r$_{helio}$ & $\Delta_{obs}$ & $\alpha$ & OD/ &  repeti- & duration \\
mid-time    & [$\mu$m]        & [mJy]  & [mJy]         &  [AU]      & [AU]            & [deg]  & OBSID & tions & [s]   \\
        \noalign{\smallskip}
        \hline
        \noalign{\smallskip}
\multicolumn{10}{l}{first visit\tablefootmark{a} on Jan.\ 6, 2013:} \\
        \noalign{\smallskip}
%
2456298.50745 &  70.0 & 36.3 &  1.1 &  1.03593 & 0.096247 & +60.44 & 1333/1342258557 & 1- 6 & 1928 \\   
2456298.50745 & 160.0 &  8.7 &  3.3 &  1.03593 & 0.096247 & +60.44 & 1333/1342258557 & 1- 6 & 1928 \\   
2456298.53059 & 100.0 & 22.8 &  1.7 &  1.03599 & 0.096234 & +60.40 & 1333/1342258558 & 1- 7 & 2012 \\   
2456298.53059 & 160.0 &  7.4 &  3.8 &  1.03599 & 0.096234 & +60.40 & 1333/1342258558 & 1- 7 & 2012 \\   
2456298.55258 &  70.0 & 37.5 &  1.3 &  1.03604 & 0.096221 & +60.36 & 1333/1342258559 & 1- 6 & 1730 \\   
2456298.55258 & 160.0 &  9.8 &  2.5 &  1.03604 & 0.096221 & +60.36 & 1333/1342258559 & 1- 6 & 1730 \\   
2456298.57455 & 100.0 & 25.0 &  1.5 &  1.03609 & 0.096208 & +60.32 & 1333/1342258560 & 1- 7 & 2012 \\   
2456298.57455 & 160.0 &  8.2 &  2.2 &  1.03609 & 0.096208 & +60.32 & 1333/1342258560 & 1- 7 & 2012 \\   
\noalign{\smallskip}
\multicolumn{10}{l}{combined first visit:} \\
\noalign{\smallskip}
2456298.53194 &  70.0 & 36.08 &  0.92 & 1.03599 & 0.096233 & +60.40 & 1342258557 + 59   & all & 3658 \\ 
2456298.55394 & 100.0 & 22.56 &  1.17 & 1.03604 & 0.096220 & +60.36 & 1342258558 + 60   & all & 4024 \\ 
2456298.54375 & 160.0 &  9.41 &  1.29 & 1.03602 & 0.096226 & +60.37 & 1342258557 ... 60 & all & 7682 \\ 
\noalign{\smallskip}
\multicolumn{10}{l}{second visit\tablefootmark{b} on Mar.\ 14, 2013:} \\
\noalign{\smallskip}
%
2456365.77802 &  70.0 &  \multicolumn{1}{l}{12.6\tablefootmark{c}}  & 2.7   &  1.093010 & 0.232276 & -61.38 & 1400/1342267456 &  1- 3 & 828 \\    
2456365.78760 &  70.0 &  \multicolumn{1}{l}{11.4}                   & 2.7   &  1.093003 & 0.232307 & -61.38 & 1400/1342267456 &  4- 6 & 828 \\    
2456365.79719 &  70.0 &  \multicolumn{1}{l}{10.4}                   & 2.7   &  1.092996 & 0.232338 & -61.39 & 1400/1342267456 &  7- 9 & 828 \\    
2456365.80677 &  70.0 &  \multicolumn{1}{l}{12.5\tablefootmark{c}}  & 2.6   &  1.092989 & 0.232368 & -61.39 & 1400/1342267456 & 10-12 & 828 \\    
2456365.81635 &  70.0 &  \multicolumn{1}{l}{13.3\tablefootmark{c}}  & 2.7   &  1.092983 & 0.232397 & -61.40 & 1400/1342267456 & 13-15 & 828 \\    
2456365.82594 &  70.0 &  \multicolumn{1}{l}{12.4}                   & 2.6   &  1.092976 & 0.232427 & -61.40 & 1400/1342267456 & 16-18 & 828 \\    
\noalign{\smallskip}
\multicolumn{10}{l}{combined second visit:} \\
\noalign{\smallskip}
2456365.80198 &  70.0 &  11.20 &  1.41 &  1.09299 & 0.232352 & -61.39 & 1400/1342267456 &  selected & 2484 \\     
2456365.80198 & 160.0 & $<$3.2 &  ---  &  1.09299 & 0.232352 & -61.39 & 1400/1342267456 &  all      & 4968 \\     
\noalign{\smallskip}
\hline
     \end{tabular}
\tablefoot{\tablefoottext{a}{Light-travel time is 48.0\,s};
           \tablefoottext{b}{light-travel time is 115.9\,s};
           \tablefoottext{c}{Photometry is still affected by 1-2\,mJy residuals from the background
      elimination process, not used for the final photometry on the combined
      measurement.}}
\end{table*}

\begin{table}
     \caption{The Apophis orientation for the nominal rotation model during
              the Herschel observations. The angular velocity was 5.00\,radians/day
              during the first visit and 5.02\,radians/day during the second visit
              (around the true spin axis at the given times).
              Numbers are given in the Apophis-centric frame.
     \label{tbl:obj_coord}}
     \begin{tabular}{rrrrrr}
        \noalign{\smallskip}
        \hline
        \hline
        \noalign{\smallskip}
Julian Date & \multicolumn{2}{c}{z-axis [deg]}      & rot.\ angle & \multicolumn{2}{c}{Rotation axis [deg]} \\
mid-time    & $\lambda_{ecl}$ & $\beta_{ecl}$ & $\phi_0$ [deg]    & $\lambda_{ecl}$ & $\beta_{ecl}$ \\
        \noalign{\smallskip}
        \hline
        \noalign{\smallskip}
\multicolumn{6}{l}{first visit on Jan.\ 6, 2013:} \\
        \noalign{\smallskip}
2456298.50745 &  19.9    & -58.5    &  243.6    &   234.53 & -75.51 \\
2456298.53059 &  11.5    & -57.5    &  241.2    &   234.39 & -75.99 \\
2456298.55258 &   3.7    & -56.4    &  239.3    &   234.44 & -76.45 \\
2456298.57455 & 356.4    & -55.2    &  237.8    &   234.74 & -76.92 \\
        \noalign{\smallskip}
\multicolumn{6}{l}{second visit on Mar.\ 14, 2013:} \\
        \noalign{\smallskip}
2456365.77802 & 294.3    & -72.9    &   96.2    &   233.35 & -70.32 \\
2456365.78760 & 293.1    & -72.1    &   97.7    &   232.58 & -70.53 \\
2456365.79719 & 291.9    & -71.3    &   99.2    &   231.81 & -70.76 \\
2456365.80677 & 290.6    & -70.6    &  100.6    &   231.04 & -71.01 \\
2456365.81635 & 289.3    & -69.8    &  101.9    &   230.38 & -71.25 \\
2456365.82594 & 287.9    & -69.0    &  103.2    &   229.78 & -71.49 \\
\noalign{\smallskip}
\hline
     \end{tabular}
\end{table}

Thermophysical model (TPM) techniques are very powerful in
deriving reliable sizes and albedos. In cases where enough thermal data
are available and if there is already information about the object's
shape and spin axis then this technique also allows to solve for
thermal properties of the surface (e.g., Harris \& Lagerros \cite{harris02};
M\"uller et al.\ \cite{mueller05}). Here the radiometric analysis
was done via a thermophysical model which is based on the work by Lagerros
(\cite{lagerros96}; \cite{lagerros97}; \cite{lagerros98}).
This model is frequently and successfully applied to near-Earth
asteroids (e.g., M\"uller et al.\ \cite{mueller04b}; \cite{mueller05};
\cite{mueller11}; \cite{mueller12}; \cite{mueller13}),
to main-belt asteroids (e.g., M\"uller \& Lagerros \cite{mueller98}; 
M\"uller \& Blommaert \cite{mueller04a}), and also to more distant
objects (e.g.\ Horner et al.\ \cite{horner12}; Lim et al.\ \cite{lim10}).
The TPM takes into account the true observing and illumination geometry
for each observational data point, a crucial aspect for the interpretation
of our Apophis observations which cover before- and after-opposition 
measurements\footnote{Before opposition: object is leading the Sun, positive
phase angle in Table~\ref{tbl:obspacs}; after opposition: object is trailing
the Sun, negative phase angles}.
The TPM allows to use any available convex shape model in combination with
spin-axis properties. The heat conduction into the surface is
controlled by the thermal inertia $\Gamma$, while the infrared beaming
effects are calculated via a surface roughness model, implemented as concave
spherical crater segments on the surface and parameterized by the root
mean square (r.m.s.) slope angle. We performed our radiometric analysis with
a constant emissivity of 0.9 at all wavelengths, knowing that the emissivity
can decrease beyond $\sim$200\,$\mu$m for some objects
(e.g., M\"uller \& Lagerros \cite{mueller98}; \cite{mueller02}), but 
our measurements are all at shorter wavelengths.
We used a mean absolute magnitude\footnote{The mean absolute magnitude corresponds to the mean
observed cross-section.} of H$_V$ = 19.09 $\pm$ 0.19\,mag which was derived by Pravec et
al.\ (\cite{pravec14}) under the assumption of a slope parameter of
G=0.24 $\pm$ 0.11.

\subsubsection{Initial estimate for flux change due to geometry}

The average observed Apophis flux at 70\,$\mu$m changed from 36.7\,mJy on
Jan.\ 6 to 11.2\,mJy on Mar.\ 14, 2013, resulting in a flux ratio
FD$_{epoch1}$/FD$_{epoch2}$ of 3.2. This ratio is driven by (i) the change in observing
geometry (r, $\Delta$, $\alpha$); (ii) the change in cross section due to
the object's non-spherical shape and the different rotational phase;
(iii) thermal effects which transport heat to non-illuminated parts of the surface.

Assuming a spherical object in instantaneous equilibrium with solar
insolation (thermal inertia equals zero) would produce a very different
70\,$\mu$m flux ratio FD$_{epoch1}$/FD$_{epoch2}$ of 6. This calculation
was the baseline for our planning of the Herschel observations in March
(second epoch measurement) where we expected to see approximately 6\,mJy instead
of the observed 11.2\,mJy.
The discrepancy between expectations and observations shows that
changes in the observed cross section and thermal effects, in addition to
the changes in observing geometry, play a significant role and are key
elements for our radiometric analysis.

\subsubsection{Initial estimate for flux change due to shape effects}

With the availability of Apophis' shape model and rotational properties
it is also possible to calculate the influence of the apparent
cross section on the observed flux. Apophis was showing a 1.21 times
larger cross section during the second epoch as compared to the first epoch.
The combined geometry and cross-section change would result in 
a 70\,$\mu$m flux ratio FD$_{epoch1}$/FD$_{epoch2}$ of about 3.7 which
is still significantly larger than the observed ratio of 3.2.
This is a strong indication that thermal effects play an important
role. The effect can also nicely be seen in
Figure~\ref{fig:tpm}: before opposition we see the object under a
phase angle of about +60$^{\circ}$ with a cold morning terminator,
while in the second epoch we have seen Apophis at about -61$^{\circ}$
with a warm evening side which has just rotated out of the Sun.
In both cases thermal effects play a strong role: during the
first epoch a substantial part of the surface heat is transported to
the non-visible side while in the second epoch the heat transport
to the non-illuminated part remains visible.


\subsubsection{Radiometric analysis of the first epoch data}
\label{sec:ep1}

\begin{figure}[h!tb]
  \rotatebox{90}{\resizebox{!}{\hsize}{\includegraphics{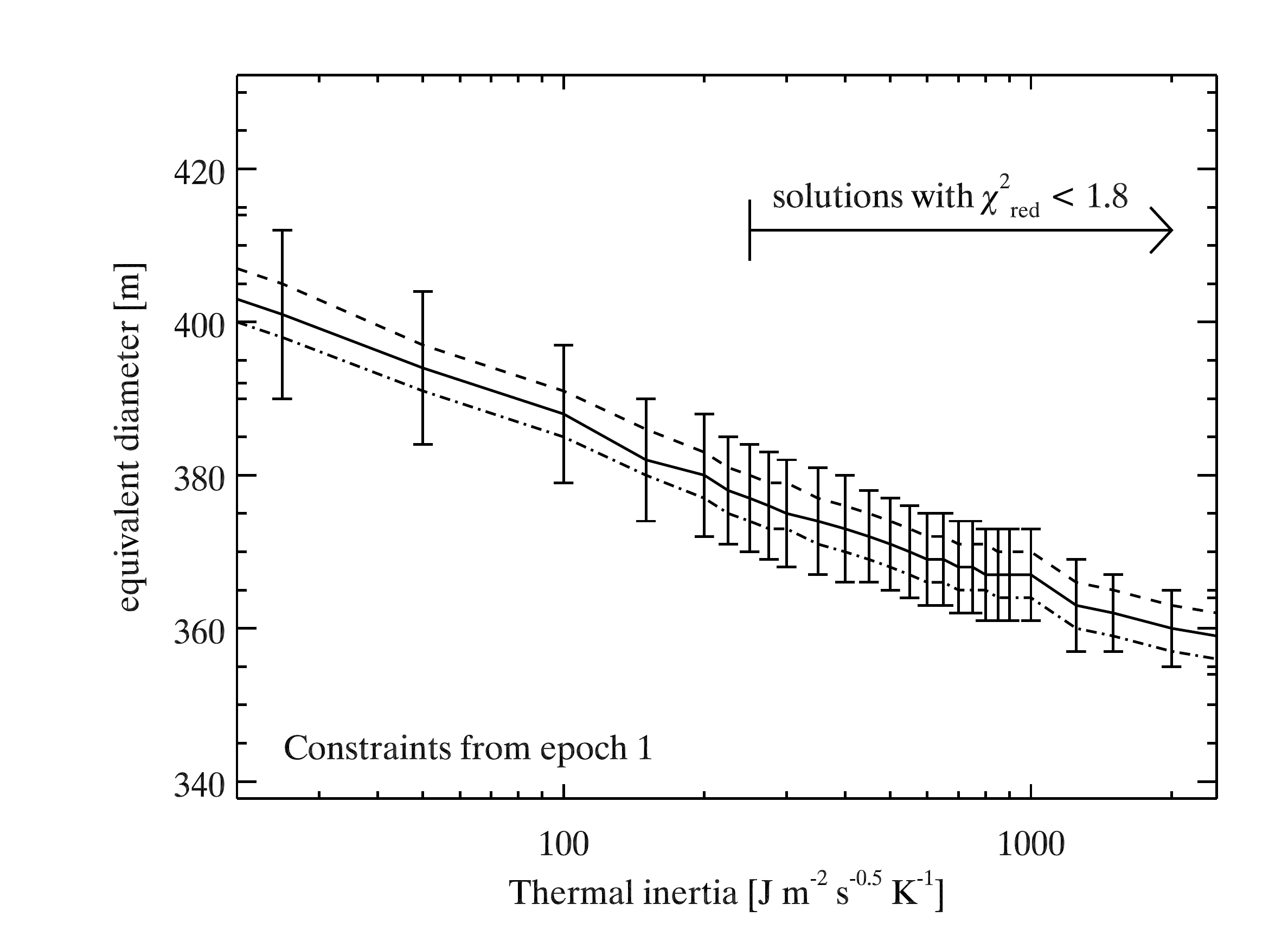}}}
  \rotatebox{90}{\resizebox{!}{\hsize}{\includegraphics{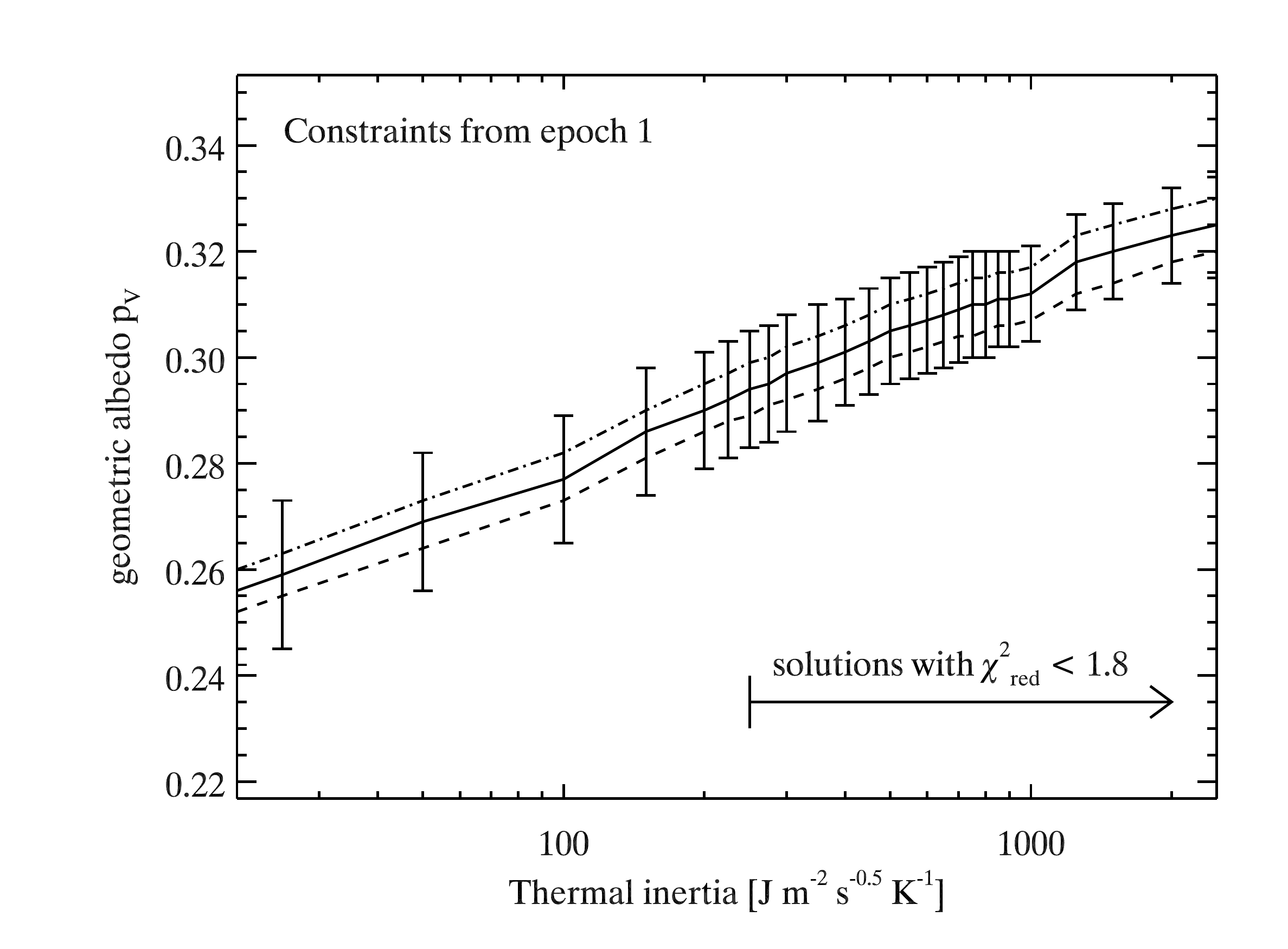}}}
  \caption{The radiometrically derived size (top) and albedo (bottom)
           as a function of thermal inertia. The influence of model
           surface roughness is shown as dashed (low roughness) and
           dotted-dashed (high roughness) lines. The errorbars indicate
           the standard deviation of observation-to-model ratios
           for our epoch-1 measurements.
     \label{fig:DpV}}
\end{figure}

\begin{figure}[h!tb]
  \rotatebox{90}{\resizebox{!}{\hsize}{\includegraphics{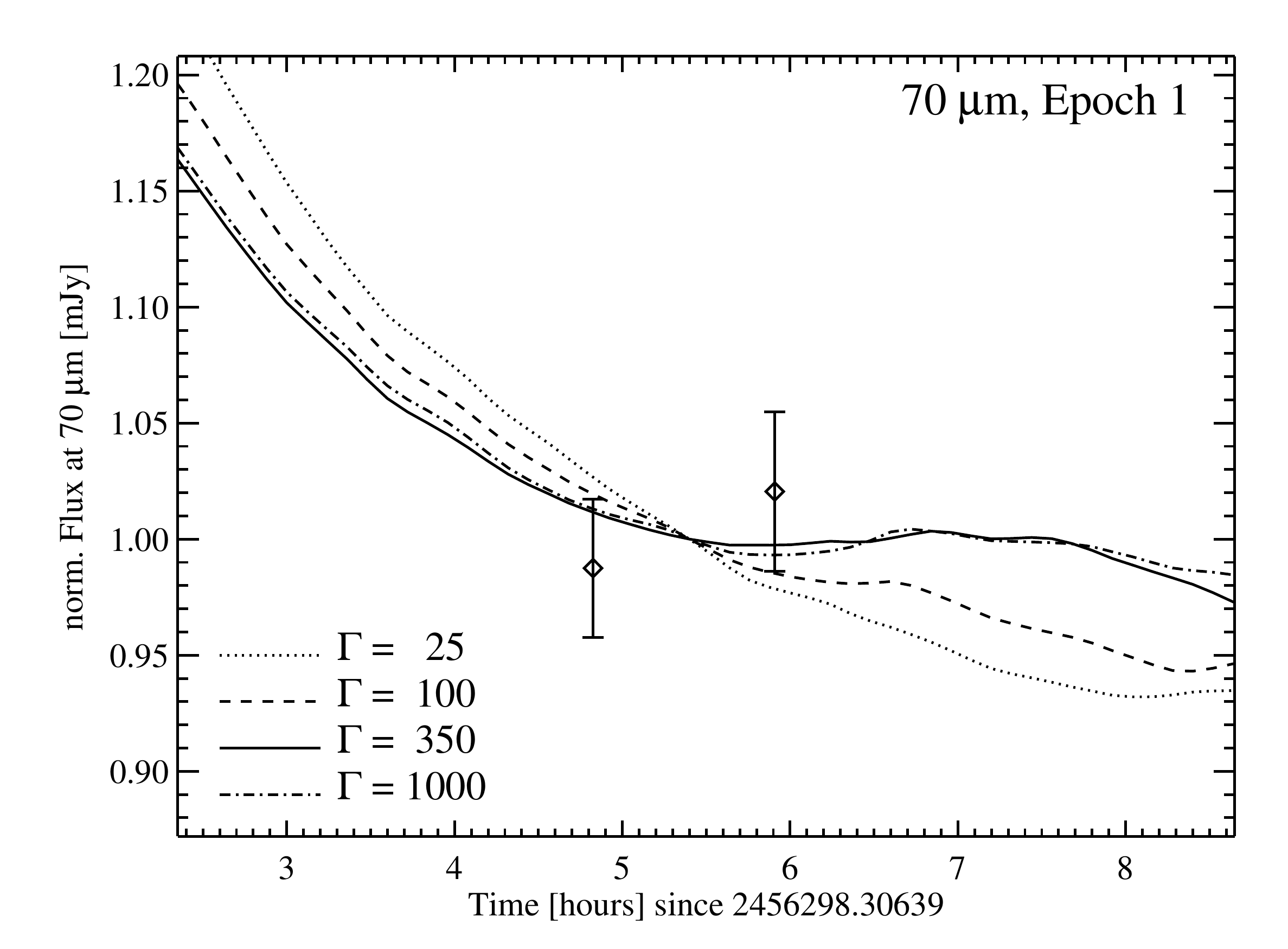}}}
  \rotatebox{90}{\resizebox{!}{\hsize}{\includegraphics{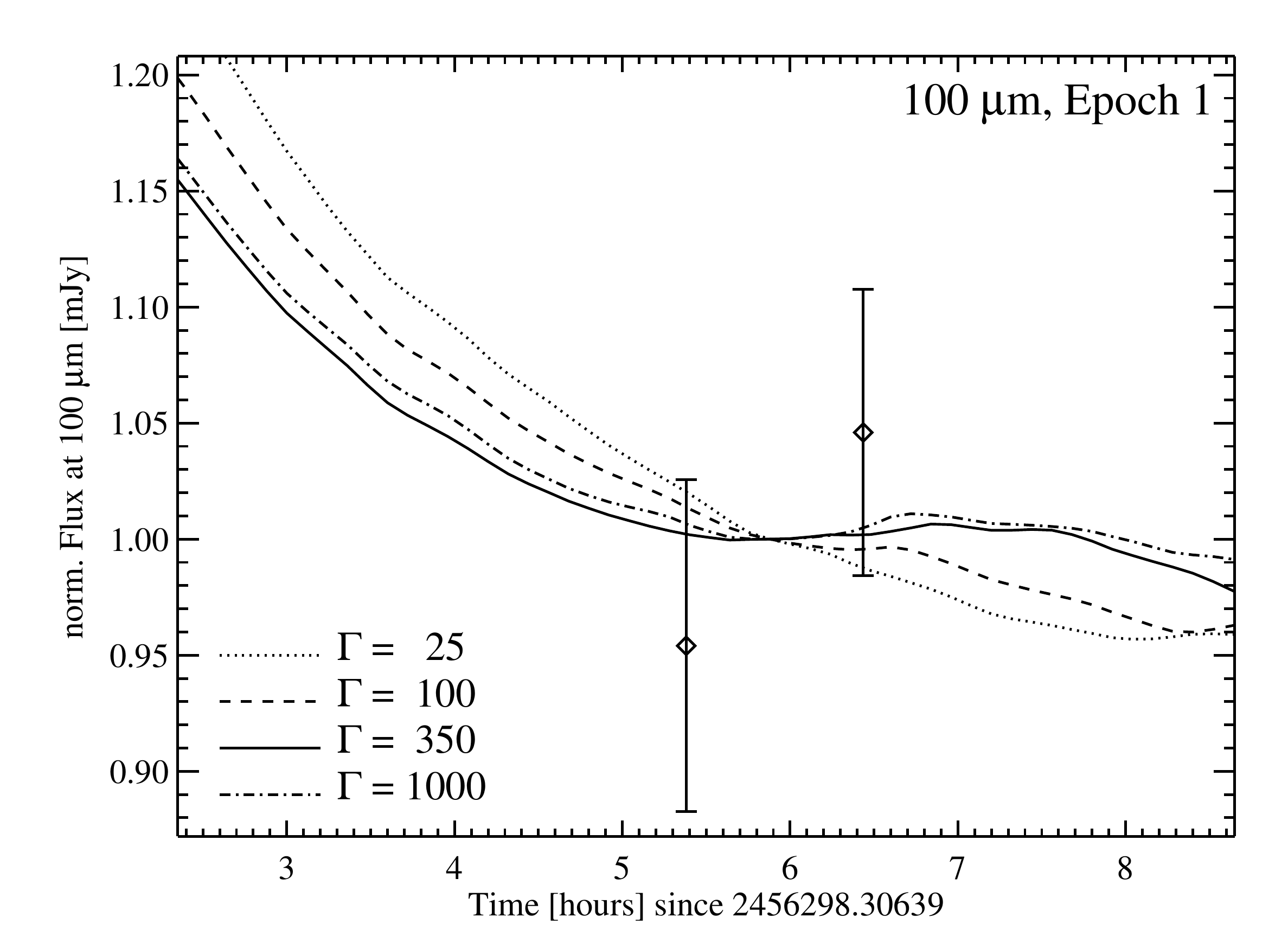}}}
  \caption{The TPM lightcurves at 70\,$\mu$m (top) and
           at 100\,$\mu$m (bottom) together with the observed
           fluxes and their errorbars, all normalised at mid-time.
           The influence of thermal inertia on the lightcurve
           is clearly visible and the measurements seem to follow
           the higher-inertia curves.
     \label{fig:TLC}}
\end{figure}

Looking at Figure~\ref{fig:tpm} (left side), we find that the observed
flux from the first epoch data taken on Jan.\ 6, 2013 is dominated
by the illuminated/heated part of the surface and the cold morning
side does not contribute in a significant manner. However, depending on
the thermal inertia of the top-surface layer there is some heat
transported to the non-visible rear side. The conversion of the observed flux
into a size and albedo solution depends therefore on the thermal inertia
and larger values for the thermal inertia lead to smaller size
and larger albedo estimates (see Fig.~\ref{fig:DpV}). We applied the
radiometric method to all epoch-1 data (see first part of Table~\ref{tbl:obspacs})
simultaneously and derived the size (of an equal-volume sphere)
and the geometric albedo (in V-band). For the calculations we used the
true, Herschel-centric observing geometry together with the correct orientation
of the object at the time of the measurements
(see Table~\ref{tbl:obj_coord} and Fig.~\ref{fig:tpm}, left side).
For signal-to-noise reasons we used the combined 100\,$\mu$m flux
(S/N=19.3), the combined 160\,$\mu$m flux (S/N=7.3), and both
individual 70\,$\mu$m fluxes (S/N = 33.0 and 28.8).
We find acceptable size-albedo solutions\footnote{Good fit solutions in the
sense of a weighted least-squares para\-meter estimation require
$\chi^2_{reduced}$ $\lesssim$1.8 for a fit to four observational data points.}
for a wide range of thermal inertias and surface roughness settings.
Only solutions connected to thermal inertias below $\sim$250\,\,Jm$^{-2}$s$^{-0.5}$K$^{-1}$
can be excluded due to high $\chi^2$-values above 1.8. Figure~\ref{fig:DpV} (top) shows
the derived size\footnote{The size of an equal-volume sphere.} and geometric albedo
values for the full range of thermal inertias. Larger values for the
thermal inertia cause more heat transport to the non-visible rear side and
require therefore smaller sizes to explain the observed fluxes. For the albedo
there is an opposite effect and larger thermal inertias are connected to
higher albedo values. The minor influence of roughness is shown by the dashed
(low r.m.s.\ slope angle of 0.2) and dotted-dashed (high r.m.s.\ slope angle of 0.9) lines.
The errorbars indicate the standard deviations at each thermal inertia for the
size and albedo values derived from each of the four individual flux measurements.
These errorbars indicate the reproducibility of the result: the sizes and albedos
connected to each of the independent measurements are inside the shown errorbars.
The 5\% absolute calibration error of the PACS photometery (Balog et al.\ \cite{balog14})
is considered later in the discussion section (Section~\ref{sec:dis}).

The thermal inertia changes the shape of the far-IR lightcurve
considerably at the time of our observations (see Figure~\ref{fig:TLC}).
At 70\,$\mu$m (top) and at 100\,$\mu$m (bottom) there is a flat part
or even a secondary maximum developing for the higher thermal inertias.
The low thermal inertia lightcurve shows a steady decrease in flux during
the two hours of Herschel measurments. This is not seen in our time-separated
observations at 70\,$\mu$m and at 100\,$\mu$m. The completely independent
measurements in both bands seem to follow the curves for the higher thermal
inertia values. At 160\,$\mu$m the errorbars are too large to see a
similar trend. The repeated 3-band high-SNR measurements from Jan.\ 6, 2013
are therefore best fit by an object with a size of 355 to 385\,m
(the diameter of a sphere with the volume equal to the asteroid),
a geometric albedo of 0.28 to 0.33, and a thermal inertia larger than
250\,Jm$^{-2}$s$^{-0.5}$K$^{-1}$.

\subsubsection{Radiometric analysis of the second epoch data}
\label{sec:ep2}

Figure~\ref{fig:tpm} (right side) illustrates nicely that the observed flux
is influenced by the non-illuminated, but still warm part of the
surface which just rotated out of the Sun. The thermal inertia
influences the temperature distribution on the surface and very
little heat is transported to the non-visible rear side.
The conversion of the observed flux into a size and albedo solution
depends therefore much less on the thermal inertia. But here
we are encountering some problems: (1) The SNR of the second-epoch
measurement is much lower due to the 2.4 times larger Herschel-centric
distance and a significant background contamination which could not be
eliminated entirely (see Table~\ref{tbl:obspacs}). (2) We only obtained
a single-band detection at 70\,$\mu$m and an upper limit at 160\,$\mu$m,
but no 100\,$\mu$m point was taken. (3) The coverage in
optical photometric points around epoch~2 is much poorer resulting in
a less accurate model at the given orientation (see Pravec
et al.\ \cite{pravec14}). The synthetic model
lightcurve of the best-fit solution shows a local maximum on the
decreasing branch and the reliability of the calculated cross-section
is not clear.

We calculated for each thermal inertia the radiometric size and albedo
solutions together with the corresponding uncertainty range.
The 13\% error in the observed 70\,$\mu$m flux translates into a
6\% error in diameter and 12\% error in albedo, the 160\,$\mu$m detection
limit unfortunately does not constrain the solution in a noticable way.
As a consequence, the full range of thermal inertias is compatible
with our epoch~2 data. The corresponding size and albedo values range
from 370\,m to 430\,m and from 0.30 to 0.22, respectively.

\subsubsection{Radiometric analysis of the combined data set}

\begin{figure}[h!tb]
  \rotatebox{90}{\resizebox{!}{\hsize}{\includegraphics{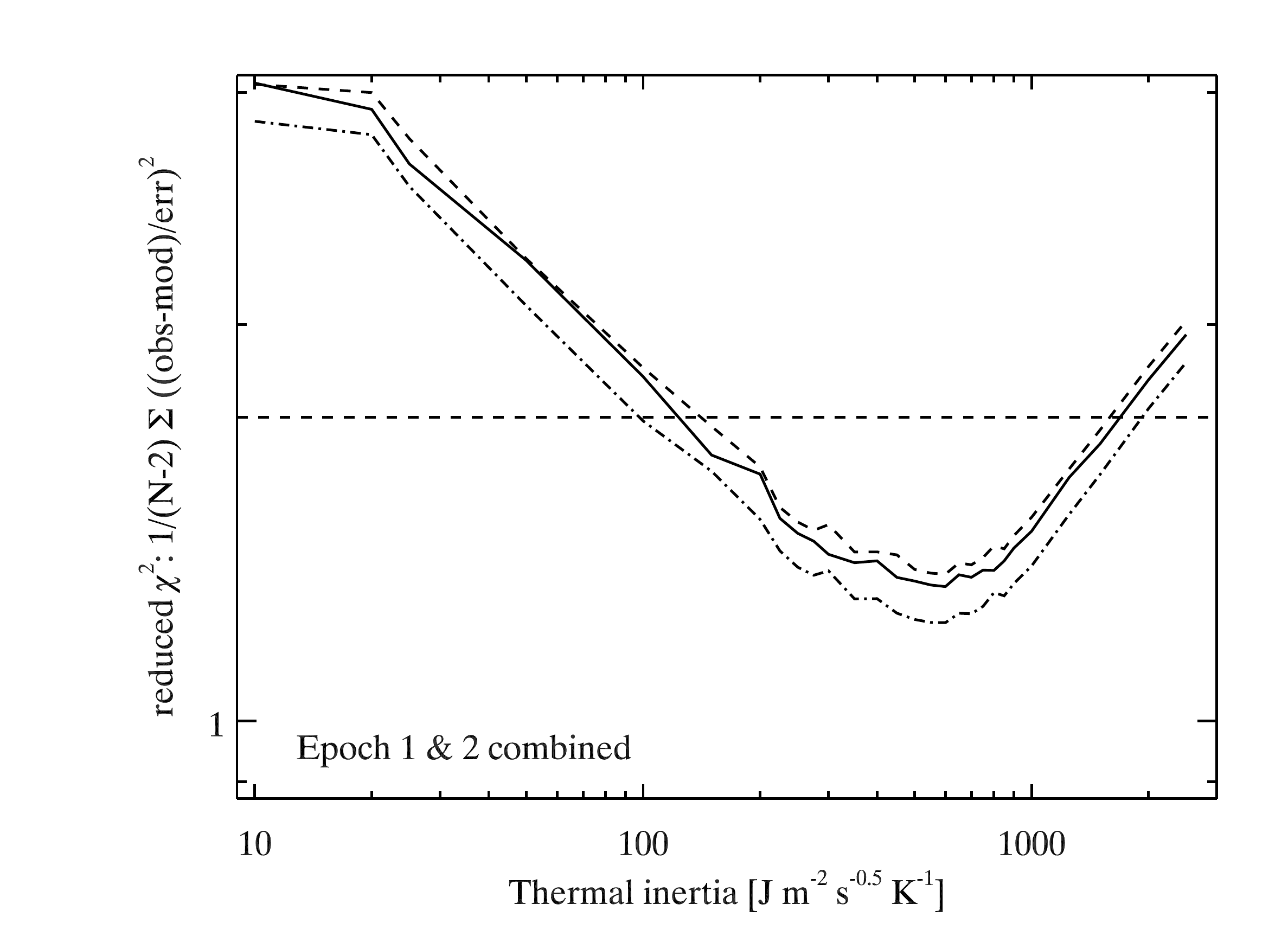}}}
  \caption{Reduced $\chi^2$-values calculated for the radiometric analysis
           of the combined epoch-1 and epoch-2 dataset. The dashed line shows the low
           roughness case, while the dashed-dotted line represents the
           very high roughness case. Good-fit solutions are found below
           the dashed horizontal line representing the reduced $\chi^2$
           threshold for five measurements at 1.7.
     \label{fig:chi2_test}}
\end{figure}

As a final step, we combined the radiometric results of Sections~\ref{sec:ep1}
and \ref{sec:ep2} while considering the derived errors. We calculated for each
thermal inertia the weighted mean size and albedo solution and used our
TPM setup (considering also the changing orientation state of the object)
to make flux predictions for the four epoch-1 and one epoch-2 data
points. Figure~\ref{fig:chi2_test} shows the reduced $\chi^2$ values together
with the 1-$\sigma$ confidence level for five independent measurements which
is around 1.7. The three different levels of surface roughness are shown
as dashed line ($\rho$ = 0.2, low roughness), solid line ($\rho$ = 0.5, intermediate
roughness), and dashed-dotted line ($\rho$ = 0.9, very high roughness).
The best solution is found at thermal inertia values around 600\,Jm$^{-2}$s$^{-0.5}$K$^{-1}$,
which is about mid-way inside the $\sim$100-1500\,Jm$^{-2}$s$^{-0.5}$K$^{-1}$
formal acceptance range. The connected size and albedo values are 368-374\,m and
0.30-0.31, respectively, with this solution being dominated by the high-quality
epoch-1 data. Giving a stronger weight to the epoch-2 observations shifts the
$\chi^2$-minima to lower thermal inertias:
if we weight the epoch-1 and epoch-2 solutions simply by the number of independent
measurements (here 4:1) then we find the $\chi^2$-minima at a thermal inertia of
around 300-350\,Jm$^{-2}$s$^{-0.5}$K$^{-1}$ and values above
800\,Jm$^{-2}$s$^{-0.5}$K$^{-1}$ would be excluded. The corresponding sizes
are about 10\,m larger and the albedo is around 0.29, but the overall match to
the observations is degraded with reduced $\chi^2$ values just below 1.7.
This kind of ``weighting by number of observations" is somewhat arbitrary,
but it shows how a better balanced (higher S/N) second epoch measurement
could have influenced our results. In Section~\ref{sec:dis} we continue with
the correct weighting of the observations taking into account the observational errorbars.

\section{Discussions}
\label{sec:dis}

The radiometric method has been found to work reliably for objects where
shape and spin properties are known (e.g., O'Rourke et al.\ \cite{orourke12}
or M\"uller et al.\ \cite{mueller14}). The application to tumbling objects
is more complex and requires the knowledge of the object's orientation
and its spin axis at the times of the thermal measurements. For our epoch-1
data set, the tumbling is not critical since the observed flux is clearly
dominated by the illuminated part of the surface. The observed flux is not
influenced by the path of the heat transport to the non-visible rear side,
independent whether the object rotates around the moment of inertia or the
true spin axis.
For our epoch-2 data, the situation is slightly different since
the temperature distribution on the warm evening side contributes to
the observed disk-integrated flux. In this case the tumbling causes a slight
spatial displacement of the contributing warm region close to the terminator.
It may be that our epoch-2 flux is slightly influenced by this effect and
that our model predictions are therefore too low. A careful investigation
showed us that the temperature of a very small region close to the rim
and outside the direct sun illumination might in reality be higher
than in our TPM calculations. But the impact on the disk-integrated
flux is well below 5\% and the consequences for our radiometric
results are negligible. The error bars in the epoch-2 observation
are simply too large.

\begin{figure}[h!tb]
  \rotatebox{90}{\resizebox{!}{\hsize}{\includegraphics{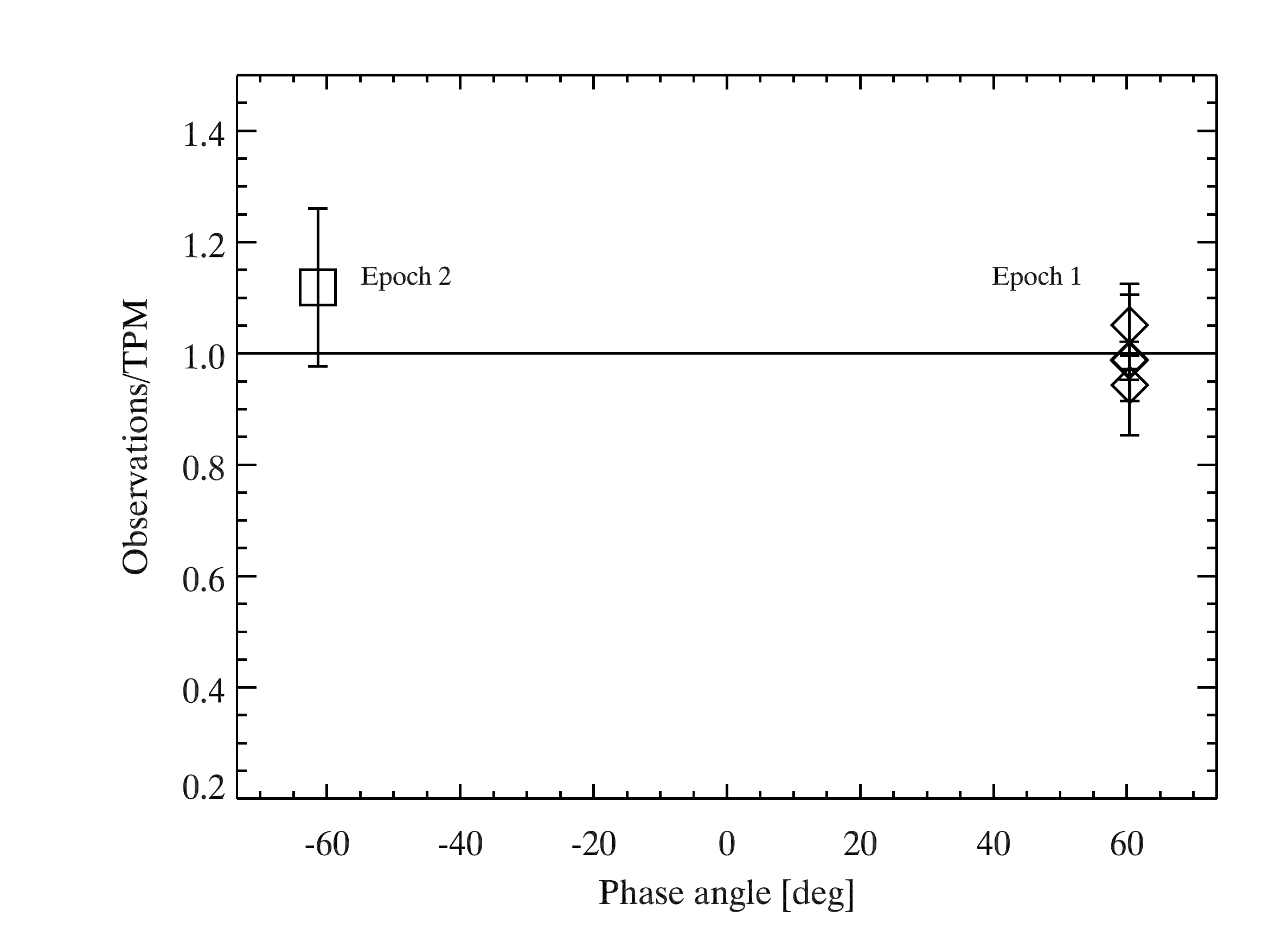}}}
  \rotatebox{90}{\resizebox{!}{\hsize}{\includegraphics{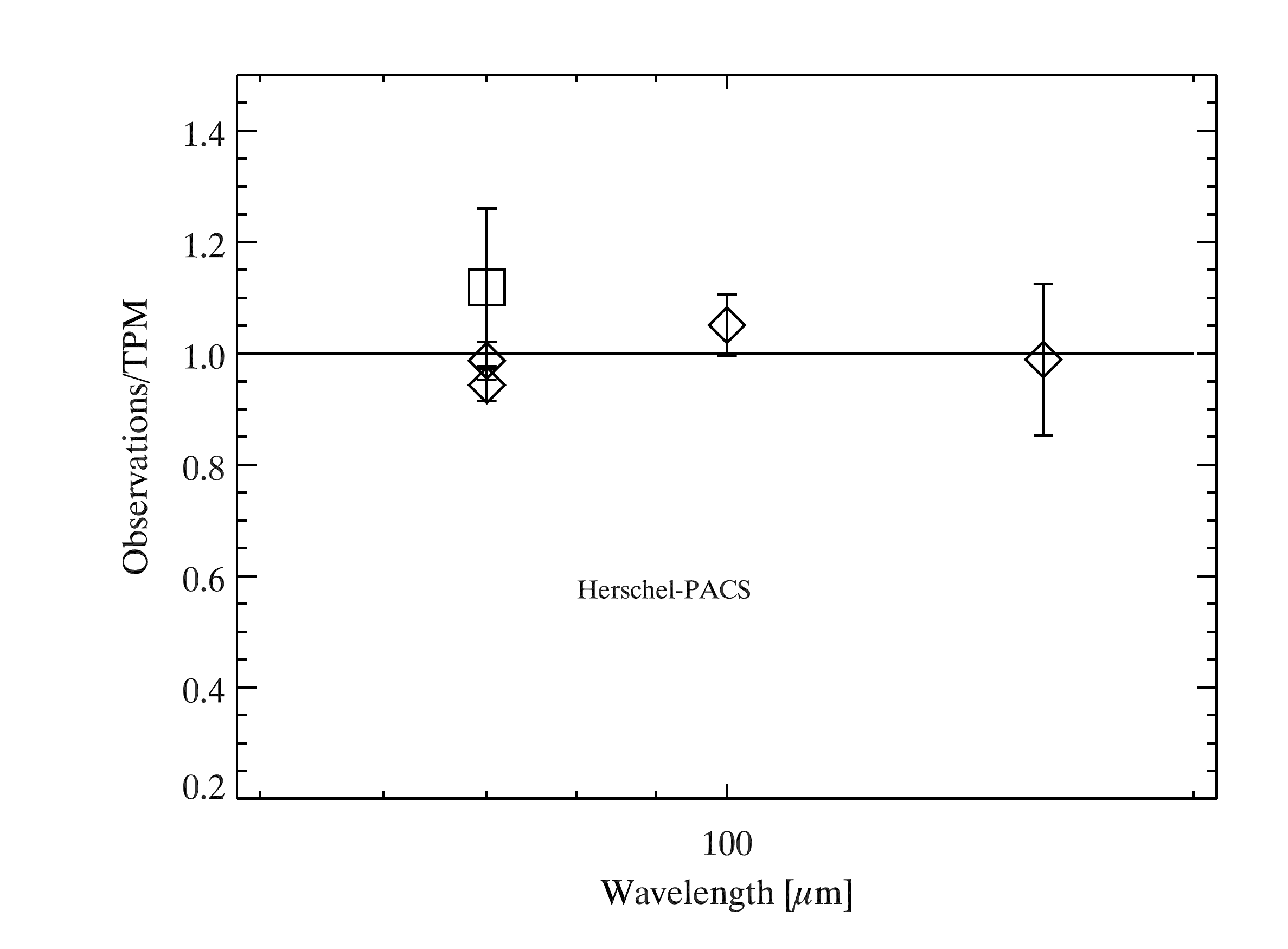}}}
  \caption{The calibrated PACS observations divided by the best TPM solution as
           a function of phase angle (top) and as a function of wavelength (bottom).
     \label{fig:obsmod}}
\end{figure}

The final uncertainties of the derived size and albedo solutions depend mainly on the quality of the thermal
measurements. A 10\% flux error translates typically in a 5\% error in
equivalent size and about 10\% in geometric albedo. With several independent
measurements the errors can reduce to even smaller values. But this is only
the case when the H-magnitude is precisely known and the thermal inertia
is well constrained by the available observational dataset.
Our dataset has a good coverage in thermal wavelengths, as well as
phase angles before and after opposition which is sufficient to determine
the thermal inertia reliably. However, due to the above mentioned problems
with epoch~2 the situation is not perfect. The epoch-1 data
indicate thermal inertias larger than about 250\,Jm$^{-2}$s$^{-0.5}$K$^{-1}$,
while the combined data set excludes only the largest values above about
1500\,Jm$^{-2}$s$^{-0.5}$K$^{-1}$. Delbo et al.\ (\cite{delbo07b}) found
an average thermal inertia of 200 $\pm$ 40\,Jm$^{-2}$s$^{-0.5}$K$^{-1}$
for a sample of km-sized near-Earth objects with a maximum derived value
of 750\,Jm$^{-2}$s$^{-0.5}$K$^{-1}$. We investigated the effects of very
high thermal inertia values above 800\,Jm$^{-2}$s$^{-0.5}$K$^{-1}$ in
the context of phase-angle and wavelength trends (as shown in
Fig.~\ref{fig:obsmod}). Although statistically still possible, these high
values produce a trend in the observation-to-model ratios with phase angle
and cause also a poor match to our most reliable 70\,$\mu$m fluxes.
The most likely range for the thermal inertia is therefore
250-800\,Jm$^{-2}$s$^{-0.5}$K$^{-1}$, with our best solution connected
to 600\,Jm$^{-2}$s$^{-0.5}$K$^{-1}$. These high values for the
thermal inertia can be explained by a mixture of (very little) low
conductivity fine regolith with larger rocks and boulders of high
thermal inertia on the surface (see also discussions in M\"uller et al.\
\cite{mueller12}, \cite{mueller13}, \cite{mueller14}).
If we take our best solution for the thermal inertia and assume a
surface density of lunar regolith (1.4\,g\,cm$^{-3}$), and that
the heat capacity is somewhere between lunar regolith (640\,J\,kg$^{-1}$K$^{-1}$)
and granite (890\,J\,kg$^{-1}$K$^{-1}$) then the thermal conductivity
$\kappa$ would be 0.3-0.4\,W\,K$^{-1}$m$^{-1}$. This is compatible
with Itokawa's 0.3\,W\,K$^{-1}$m$^{-1}$ (M\"uller et al.\ \cite{mueller05})
whereas the typical value for near-Earth asteroids is 0.08\,W\,K$^{-1}$m$^{-1}$
(Mueller M.\ \cite{mueller07}). If we take the full range of uncertainties into
account ($\Gamma$ = 250-800\,Jm$^{-2}$s$^{-0.5}$K$^{-1}$, heat capacity
450-1200\,J\,kg$^{-1}$K$^{-1}$, and surface density 1.3-2.0\,g\,cm$^{-3}$)
then the range for thermal conductivity would be 0.03-1.1\,W\,K$^{-1}$m$^{-1}$,
which is a range of two orders of magnitude.

The size range corresponding to our thermal inertia solution is
371 to 385\,m (best solution 375\,m) with a statistical error
of about 6\,m only. The smallest radiometric size solutions are
produced by the high-roughness and high-inertia settings in the TPM,
while the largest sizes are related to low-roughness/low-inertia
settings (see also Rozitis \& Green \cite{rozitis11} for a discussion
on the degeneracy between roughness and thermal inertia). Since the
PACS photometric system is only accurate on a
5\% level (Balog et al.\ \cite{balog14}), we have to consider it
also in the context of our size solution\footnote{We added quadratically
the statistical size error with a 2.5\% size error related to the
5\% in absolute flux calibration.}. The final size value is
therefore 375$^{+14}_{-10}$\,m.

Our derived albedo range of 0.28 to 0.31 (larger values for high-roughness,
high-inertia case) has a very small statistical error below 3\%. But here
we have to include the influence of the absolute flux calibration (5\%),
as well as the H-magnitude error of $\pm$ 0.19\,mag which is the dominating
factor for the final solution. Overall, we find a geometric albedo solution of
0.30$^{+0.05}_{-0.06}$. This value is in nice agreement with the Delbo et al.\
(\cite{delbo07a}) of 0.33 $\pm$ 0.08, derived from polarimetric observations.
The small size solution of 270 $\pm$ 60\,m by Delbo et al.\ (\cite{delbo07a})
was mainly related to their H-magnitude which is very different from the
value by Pravec et al.\ (\cite{pravec14}) which we used here.
We can now also determine the bolometric Bond albedo A.
The uncertainty in G translates into an uncertainty in the phase
integral  q (Bowell et al.\ \cite{bowell89}), combined with a 5\%
accuracy of the q-G relation (Muinonen et al. \cite{muinonen10}),
we obtain a Bond albedo of A = q\,$\cdot$\,$p_V$ = 0.14$^{+0.03}_{-0.04}$.

\begin{figure}[h!tb]
  \rotatebox{90}{\resizebox{!}{\hsize}{\includegraphics{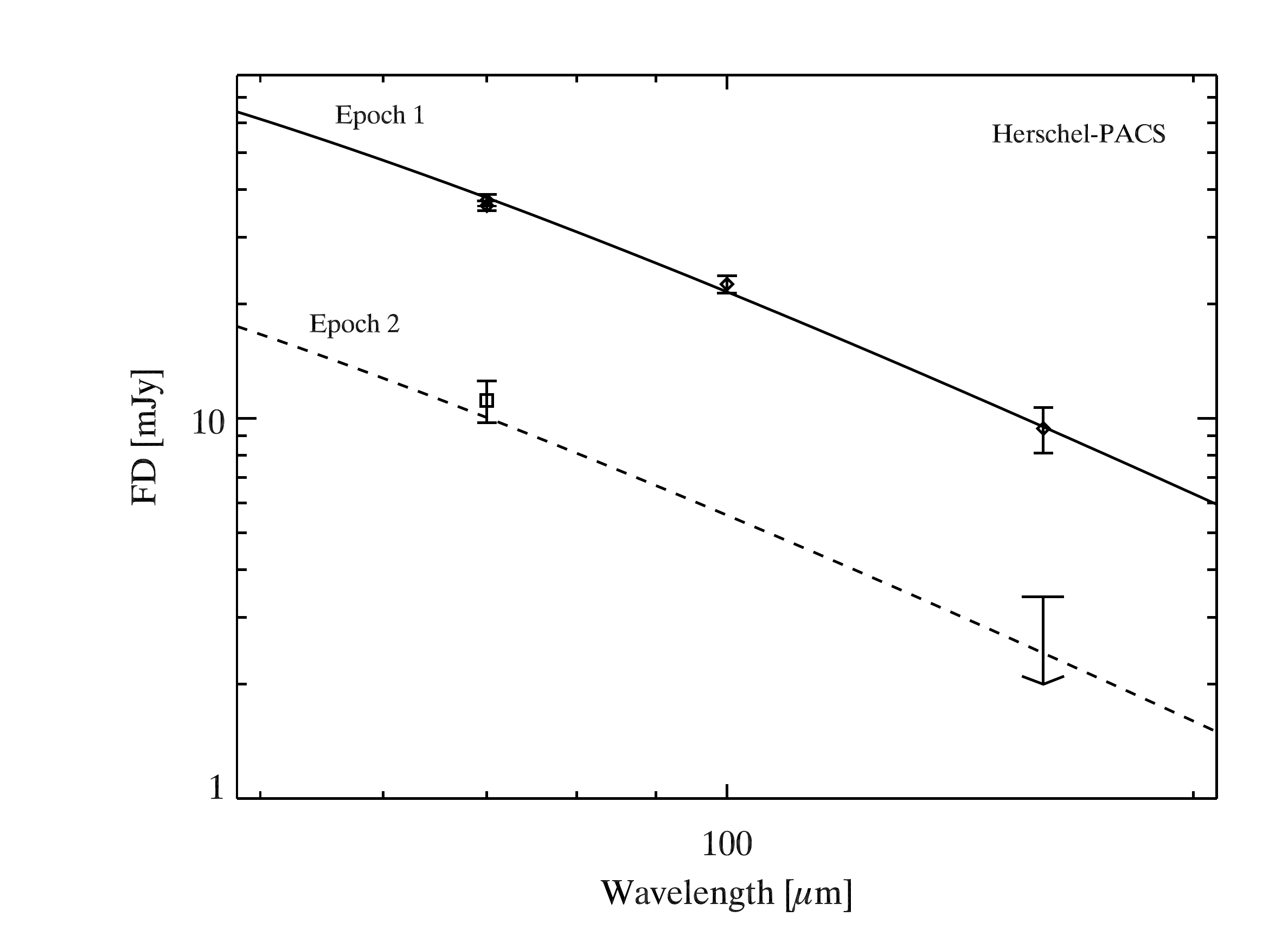}}}
  \caption{The observed absolute fluxes and the corresponding
           TPM predictions. The best TPM solution is shown as solid line
           (epoch 1) and as dashed line (epoch 2).
     \label{fig:sed}}
\end{figure}

Figures~\ref{fig:obsmod} and \ref{fig:sed} show our best model solution
at intermediate roughness level in different representations.
In Figure~\ref{fig:obsmod} we present the observations divided by
the corresponding model solutions as a function of phase angle (top)
and as a function of wavelength (bottom). No trends with phase angle
or wavelength can be seen. Figure~\ref{fig:sed} shows the observations
and the model solution on an absolute scale. Here we also show the
160\,$\mu$m upper limit from epoch 2 which is in nice agreement
with the model solution.

Binzel et al.\ (\cite{binzel09}) found compositional similarities
between 99942~Apophis and 25143~Itokawa. They both are in a similar
size range, have similar albedos and similar thermal inertias.
The measured density of Itokawa is 1.9 $\pm$ 0.13 g/cm$^3$
(Fujiwara et al.\ \cite{fujiwara06}; Abe et al.\ \cite{abe06} found
a slightly higher density of 1.95 $\pm$ 0.14\,g/cm$^3$).
Using Itokawa's density and our new size estimate gives a mass
estimate of 5.2$^{+0.7}_{-0.6}$ $\cdot$ 10$^{10}$\,kg.
Both Itokawa and Apophis have been interpreted to be analoguous
to LL chondrite meteorites (Fujiwara et al.\ \cite{fujiwara06};
Binzel et al.\ \cite{binzel09}). The bulk density of meteorites
of that type is 3.21 $\pm$ 0.22\,g/cm$^3$ (Britt \& Consolmagno
\cite{britt03}). A larger uncertainty is in the macro-porosity
of Apophis. Britt et al.\ (\cite{britt02}) report that asteroids'
macro-porosities may be up to 50\%. The porosity of Itokawa is 41\%
(Abe et al.\ \cite{abe06}). Assuming a porosity range of 30-50\%
for Apophis implies a mass between 4.4 and 6.2 $\cdot$ 10$^{10}$\,kg.

The comparison with Itokawa is interesting in many aspects: 
The rubble-pile near-Earth asteroid 25143~Itokawa has an effective
size of 327.5 $\pm$ 5.5\,m (Fujiwara et al.\ \cite{fujiwara06}),
just 13\% smaller than Apophis. Both objects have almost identical
geometric albedos: 0.29 $\pm$ 0.02 for Itokawa (Bernardi et al.\
\cite{bernardi09}) compared to 0.30$^{+0.05}_{-0.06}$ for Apophis.
Also the thermal inertias compare very well: M\"uller et al.\
(\cite{mueller14}) found 700 $\pm$ 200\,Jm$^{-2}$s$^{-0.5}$K$^{-1}$ for Itokawa,
well within the derived range for Apophis.
Itokawa has a SIV-type taxonomic classification (Binzel et al.\ 
\cite{binzel01}) and the Hayabusa data revealed an olivine-rich
mineral assemblage silimar to LL5 or LL6 chondrites (Abe et al.\
\cite{abe06}; Okada et al.\ \cite{okada06}). Apophis is characterised
as an Sq-type that most closely resembles LL ordinary chondrite meteorites
(Binzel et al.\ \cite{binzel09}).
The high thermal inertia indicates a lack (or only very small amounts)
of low-conductivity fine regolith on the surface. The formation
of a thick regolith (typically with
$\Gamma$-values below 100\,Jm$^{-2}$s$^{-0.5}$K$^{-1}$) might have
been hampered by frequent seismic influences. Such processes can
reorganise the body's interior and surface over short time scales
if the object has a rubble-pile structure.
Apophis is also in the size range predominated by asteroids
with cohesionless structures (Pravec et al.\ \cite{pravec07}).
On the other hand, the density of S-type asteroids is distributed
in a very narrow density interval, slightly below the density
of their associated meteorites, the ordinary chondrites (Carry
\cite{carry12}). The macroporosity for this type of asteroids
is generally smaller than 30\% and pointing to coherent interiors,
with cracks and fractures, but not rubble piles. Interestingly,
the four S-type asteroids listed by Carry (\cite{carry12}) with
sizes below a few kilometres and with high quality density
information (quality codes A, B, or C) all have densities
below 2\,g\,cm$^{-3}$ and a porosity of 40\% or above, indicative
of a rubble-pile structure. Overall, Apophis' size, the surface characteristics related
to a relatively high thermal inertia, and the comparison with similar-size objects,
make a cohesionless structure more likely.

%
%
%
%
%

The newly derived properties will influence the long-term
orbit predictions. Chesley et al.\ (\cite{chesley03}; \cite{chesley08})
and Vokrouhlick\'y et al.\ (\cite{vokrouhlicky08}) found that the
Yarkovsky effect which is due to the recoil of thermally re-radiated
sunlight is acting on many near-Earth asteroids. It is the most significant
non-gravitational force to be considered for risk analysis studies
(e.g., Giorgini et al.\ \cite{giorgini02}, \cite{giorgini08}; Chesley
\cite{chesley06}). The calculation of the Yarkovsky orbit drift
requires -in addition to the spin state which was determined by Pravec
et al.\ (\cite{pravec14})- also some knowledge about the object's size,
bulk density, and surface thermal inertia.
Our work will contribute with information about size
and thermal inertia (Vokrouhlick\'y et al., \cite{vokrouhlicky14}).
The bulk density can be estimated from the Yarkovsky-related orbit
change, expected to be detected by radar observations during the
next close Earth approach in September 2021 (Farnocchia et al.\ \cite{farnocchia13}).
\v{Z}i\v{z}ka \& Vokrouhlick\'y (\cite{zizka11}) showed that also
the solar radiation pressure has a small, but relevant effect on Apophis' orbit
which might be noticable after the very close Earth encounter in 2029.
Here it is mainly the size and bulk density which play a role.
The combined non-gravitational forces -Yarkovsky effect and solar radiation pressure-
cause small orbit drifts up to a few kilometers per decade in case of Apophis
(Farnocchia et al.\ \cite{farnocchia13}).
In comparison, the extension of the keyholes associated with Earth-impacts after the 2029 close
encounter are in the order of a 100\,m or smaller. The studies of the non-gravitational
orbit perturbations are therefore important to estimate the distance between
the true trajectory and the locations of the dangerous keyholes.

\section{Conclusions}
\label{sec:con}

The shape and spin properties of Apophis presented by Pravec et al.\ (\cite{pravec14})
were the key elements for our radiometric analysis. The interpretation of the $\sim$3.5\,h
of Herschel-PACS measurements in January and March 2013 was done using a well-tested and validated
thermophysical model. Applying the radiometric method to a tumbling object is more
complex, but it works reliably if the object's orientation and its spin axis
is known at the epochs of the thermal measurements. We found the following results:
   \begin{enumerate}
      \item The radiometric size solution is D$_{eff}$ = 375$^{+14}_{-10}$\,m; this is the
            scaling factor for the shape model presented in Pravec et al.\ (\cite{pravec14})
            and corresponds to the size of an equal volume sphere.
      \item The geometric V-band albedo was found to be p$_V$ = 0.30$^{+0.05}_{-0.06}$, almost
            identical to the value found for the Hayabusa rendezvous target 25143~Itokawa;
            the corresponding bolometric Bond albedo A is 0.14$^{+0.03}_{-0.04}$.
      \item A thermal inertia of $\Gamma$ = 600$^{+200}_{-350}$\,Jm$^{-2}$s$^{-0.5}$K$^{-1}$
            explains best our combined dataset comprising three different bands and two different
            epochs.
      \item Using either Itokawa's bulk density information or a rock density of 3.2\,g/cm$^3$
            combined with 30-50\% porosity, we calculate a mass of (5.3 $\pm$ 0.9) $\cdot$10$^{10}$\,kg
            which is 2 to 3 times higher than earlier estimates.
      \item No information about surface roughness can be derived from
            the radiometric analysis of our measurements due to the lack
            of observations at shorter wavelengths and smaller phase angles
            close to opposition. But Apophis' thermal inertia is similar to
            the value derived for Itokawa and this might point to a surface
            of comparable roughness.
      \item Apophis' size, the surface characteristics related to
            the high thermal inertia, and the comparison with similar-size
            objects, make a cohesionless structure more likely.
  \end{enumerate}


  The interior structure -rubble pile or coherent body- is relevant in
  the context of impact scenario studies. In case of a rubble-pile structure
  (which is the more likely option)
  pre-collision encounters with planets could disrupt the body by tidal
  forces while a more solid interior would leave the object intact.
  We also expect that the newly derived properties
  will affect the long-term orbit predictions of Apophis
  which is influenced by the Yarkovsky effect and in second order
  also by the solar radiation pressure. In this context, the radiometrically
  derived size and thermal inertia will play a significant role in
  risk-analysis studies beyond Apophis' close encounter with Earth
  in 2029.

\begin{acknowledgements}
  We would like to thank the Herschel operations team which
  supported the planning and scheduling of our time-constrained
  observations. Without their dedication and enthusiasm these
  measurements would not have been possible. The first-visit data
  are part of the Herschel GT1 MACH-11 project (PI: L.\ O'Rourke),
  while the second-visit data were obtained via a dedicated DDT
  project (PI: T.\ M\"uller). The work of P.S.\ and P.P.\ was
  supported by the Grant Agency of the Czech Republic, Grant P209/12/0229,
  by the Ministry of Education of the Czech Republic, Grant LG12001,
  and by Program RVO 67985815. E.V.\ was supported by German DLR project
  funding 50\,OR\,1108.
\end{acknowledgements}


\begin{thebibliography}{}
\bibitem[2006]{abe06}
   Abe, S., Mukai, T., Hirata, N.\ et al.\ 2006,
   Science, 312, 1344-1347
\bibitem[2014]{balog14}
   Balog, Z., M\"uller, T.\ G., Nielbock, M.\ et al.\ 2014,
   Experimental Astronomy, accepted, DOI: 10.1007/s10686-013-9352-3
\bibitem[2009]{bernardi09}
   Bernardi, F., Micheli, M.\ \& Tholen, D.\ J.\ 2009,
   Meteorit. Planet. Sci., 44, 1849
\bibitem[2001]{binzel01}
   Binzel, R.\ P., Rivkin, A.\ S., Bus, S.\ J.,\ et al.\ 2001,
   Meteorit. Planet. Sci. (Suppl.), 36, A20
\bibitem[2009]{binzel09}
   Binzel, R.\ P., Rivkin, A.\ S., Thomas, C.\ A.\ et al.\ 2009,
   Icarus 200, 480-485
\bibitem[1989]{bowell89}
   Bowell, E., Hapke, B., Domingue, D.\ et al.\ 1989,
   in Asteroids II, R.P.\ Binzel, T.\ Gehrels, M.\ Shapley Matthews (Eds.), Univ.\ of Arizona Press, 524-556
\bibitem[2002]{britt02}
   Britt, D.\ T., Yeomans, D., Housen, K., Consolmagno, G.\ 2002,
   in Asteroids III, W.\ F.\ Bottke Jr., A.\ Cellino, P.\ Paolicchi, and R.\ P.\ Binzel (eds),
   University of Arizona Press, Tucson, p.485-500
\bibitem[2003]{britt03}
   Britt, D.\ T.\ \& Consolmagno, G.\ 2003,
   M\&PS, 38, 1161
\bibitem[2012]{carry12}
   Carry, B.\ 2012,
   P\&SS 73, 98-118
\bibitem[2003]{chesley03}
   Chesley, S.\ R., Ostro, S.\ J., Vokrouhlick\'y, D.\ et al. 2003,
   Science 302, 1739-1742
\bibitem[2006]{chesley06}
   Chesley, S.\ R.\ 2006,
   in: Lazzaro, D., Ferraz-Mello, S., Fern\'andez, J.\ (Eds),
   Asteroids, Comets, Meteors, Cambridge University Press, Cambridge, 215-228
\bibitem[2008]{chesley08}
   Chesley, S.\ R., Vokrouhlick\'y, D., Ostro, S.\ J.\ et al.\ 2008,
   Asteroids, Comets, Meteors 2008, Baltimore, Maryland, Contribution No.\ 1405, Paper Id.\ 8330
\bibitem[2007a]{delbo07a}
   Delbo, M., Cellino, A., Tedesco, E.\ F.\ 2007,
   Icarus 188, 266-269
\bibitem[2007b]{delbo07b}
   Delbo, M., dell'Oro, A., Harris, A.\ W.\ et al.\ 2007,
   Icarus, 190, 236
\bibitem[2013]{farnocchia13}
   Farnocchia, D., Chesley, S.\ R., Chodas, P.\ W.\ et al.\ 2013,
   Icarus 224, 192-200
\bibitem[2006]{fujiwara06}
   Fujiwara, A., Kawaguchi, J., Uesugi, K.\ et al.\ 2006,
   Science, 312, 1330
\bibitem[2002]{giorgini02}
   Giorgini, J.\ D.,  .... 2002,
   Science 296, 132-136
\bibitem[2008]{giorgini08}
   Giorgini, J.\ D., Benner, L.\ A.\ M., Ostro, S.\ J.\ et al.\ 2008,
   Icarus 193, 1-19
\bibitem[2002]{harris02}
   Harris, A.\ W.\ \& Lagerros, J.\ S.\ V.\ 2002,
   in Asteroids III, W.\ F.\ Bottke Jr., A.\ Cellino, P.\ Paolicchi, and R.\ P.\ Binzel (eds),
   University of Arizona Press, Tucson, p.205-218
\bibitem[2012]{horner12}
   Horner, J., M\"uller, T.\ G., Lykawka, P.\ S.\ 2012,
   MNRAS 423, 2587-2596
\bibitem[2001]{kaasalainen01}
   Kaasalainen, M.\ \& Torppa, J.\ 2001,
   Icarus, 153, 24
\bibitem[2001a]{kaasalainen01a}
   Kaasalainen, M., Torppa, J.\ \& Muinonen, K.\ 2001,
   Icarus, 153, 37
\bibitem[1996]{lagerros96}
   Lagerros, J.\ S.\ V.\ 1996, A\&A 310, 1011
\bibitem[1997]{lagerros97}
   Lagerros, J.\ S.\ V.\ 1997, A\&A 325, 1226
\bibitem[1998]{lagerros98}
  Lagerros, J.\ S.\ V.\ 1998, A\&A 332, 1123
\bibitem[2010]{lim10}
   Lim, T.\ L., Stansberry, J., M\"uller, T.\ G.\ et al.\ 2010,
   A\&A 518, 148-152
\bibitem[1998]{mueller98}
   M\"uller, T.\ G.\ \& Lagerros, J.\ S.\ V.\ 1998,
   A\&A, 338, 340-352
\bibitem[2002]{mueller02}
   M\"uller, T.\ G.\ \& Lagerros, J.\ S.\ V.\ 1998,
   A\&A, 381, 324-339
\bibitem[2004]{mueller04a}
   M\"uller, T.\ G.\ \& Blommaert, J.\ A.\ D.\ L.\ 2004,
   A\&A, 418, 347-356
\bibitem[2004]{mueller04b}
   M\"uller, T.\ G., Sterzik, M.\ F., Sch\"utz, O.\ et al.\ 2004,
   A\&A, 424, 1075-1080
\bibitem[2007]{mueller07}
   Mueller, M.\ 2007,
   PhD thesis, Freie Universit\"{a}t Berlin, Germany, {\tt http://www.diss.fu-berlin.de/diss/receive/FUDISS\_thesis\_000000002596}
\bibitem[2005]{mueller05}
  M\"uller, T.\ G., Sekiguchi, T., Kaasalainen, M.\ et al.\ 2005,
  A\&A, 443, 347-355
\bibitem[2011]{mueller11}
   M\"uller, T.\ G., \v{D}urech, J., Hasegawa, S.\ et al.\ 2011,
   A\&A, 525, 145
\bibitem[2012]{mueller12}
  M\"uller, T.\ G., O'Rourke, L., Barucci, A.\ M.\ et al.\ 2012,
  A\&A, 548, 36-45
\bibitem[2013]{mueller13}
  M\"uller, T.\ G., Miyata, T., Kiss, C.\ et al.\ 2013,
  A\&A, 558, 97
\bibitem[2014]{mueller14}
  M\"uller, T.\ G., Hasegawa, S.\ \& Usui, F.\ 2014,
  PASJ, accepted
\bibitem[2010]{muinonen10}
  Muinonen, K., Belskaya, I.\ N., Cellino, A.\ et al.\ 2010,
  Icarus 209, 542
\bibitem[2013]{nielbock13}
   Nielbock, M., M\"uller, T.\ G., Balog, Z.\ et al.\ 2013,
   Experimental Astronomy, 36, 631
\bibitem[2006]{okada06}
   Okada, T., Shirai, K., Yamamoto, Y.,\ et al.\ 2006,
   Science, 312, 1338
\bibitem[2012]{orourke12}
   O'Rourke, L., M\"uller, T., Valtchanov, I.,\ et al.\ 2012,
   Planet. Space Sci., 66, 192
\bibitem[2010]{pilbratt10}
  Pilbratt, G.L., Riedinger, J.R., Passvogel, T.\ et al.\ 2010, A\&A, 518, L1
\bibitem[2010]{poglitsch10}
  Poglitsch, A., Waelkens, C., Geis, N.\ et al.\ 2010, A\&A, 518, L2
\bibitem[2007]{pravec07}
  Pravec, P., Harris, A.\ W., Warner, B.\ D.\ 2007,
  in: Milani, A., Valsecchi, G.\ B., Vokrouhlick\'y, D.\ (Eds.),
  Proc.\ IAU Symp, vol.\ 236. Cambridge Univ.\ Press, Cambridge, 167-176
\bibitem[2014]{pravec14}
  Pravec, P., Scheirich, P., \v{D}urech, J.\ et al.\ 2014,
  Icarus, in press
\bibitem[2011]{rozitis11}
         Rozitis, B.\ \&  Green, S.\ F.\ 2011,
         MNRAS, 415, 2042
\bibitem[2010]{scheirich10}
  Scheirich, P., Durech, J., Pravec, P.\ et al.\ 2010,
  M\&PS, 45, 1804
\bibitem[2008]{vokrouhlicky08}
  Vokrouhlick\'y, D., Chesley, S.\ R., Matson, R.\ D.\ 2008,
  AJ 135, 2336-2340
\bibitem[in preparation]{vokrouhlicky14}
  Vokrouhlick\'y, D.\ et al., in preparation
\bibitem[2013]{wlodarczyk13}
  Wlodarczyk, I.\ 2013,
  MNRAS, 434, 3055-3060
\bibitem[2011]{zizka11}
  \v{Z}i\v{z}ka, J.\ \& Vokrouhlick\'y, D.\ 2011,
  Icarus, 211, 511-518
\end{thebibliography}
\end{document}